\documentclass[aps,prb,amsmath,amssymb,floatfix,twocolumn]{revtex4-2}

\DeclareUnicodeCharacter{2212}{-}
\DeclareUnicodeCharacter{2032}{-}

\usepackage{longtable}
\usepackage{epsfig}
\usepackage{multirow}
\usepackage{bbold}
\usepackage{graphicx}
\usepackage{booktabs,makecell}

\usepackage{color}
\usepackage{hyperref}
\usepackage{filecontents}

\usepackage[normalem]{ulem}

\usepackage{times}
\usepackage{physics}

\usepackage{amstext}
\usepackage{array}

\usepackage{diagbox}

\usepackage{siunitx}

\usepackage{yfonts}

\allowdisplaybreaks

\usepackage{setspace}

\usepackage{newfloat}
\DeclareFloatingEnvironment[name={FIG. S}]{suppfigure}

\newcommand{\ncmd}{\newcommand}
\ncmd{\nn}{\nonumber}
\ncmd{\Lam}{\Lambda}
\ncmd{\lam}{\lambda}
\ncmd{\Gam}{\Gamma}
\ncmd{\gam}{\gamma}
\ncmd{\sig}{\sigma}
\ncmd{\Dl}{\Delta}
\ncmd{\dl}{\delta}
\ncmd{\kap}{\kappa}
\ncmd{\mc}{\mathcal}
\ncmd{\veps}{\varepsilon}
\ncmd{\vphi}{\varphi}
\ncmd{\vtheta}{\vartheta}
\ncmd{\bs}{\boldsymbol}
\ncmd{\pll}{\parallel}

\begin{document}

\title {Oxygen Vacancy Formation Energy in Metal Oxides: High Throughput Computational Studies and Machine Learning Predictions}

\author{Bianca Baldassarri$^{1}$}
\author{Jiangang He$^{2,3}$}
\author{Abhijith Gopakumar$^{2,4}$}
\author{Sean Griesemer$^{2}$}
\author{Adolfo J. A. Salgado-Casanova$^{2}$}
\author{Tzu-Chen Liu$^{2}$}
\author{Steven B. Torrisi$^{5}$}
\author{Chris Wolverton$^{1,2}$}
\email{c-wolverton@northwestern.edu}

\affiliation
{$^1$Graduate Program in Applied Physics, Northwestern University, Evanston, IL}
\affiliation
{$^2$ Department of Materials Science, Northwestern University, Evanston, IL}
\affiliation{$^3$Present: School of Mathematics and Physics, University of Science and Technology Beijing, Beijing}
\affiliation{$^4$Present: QuesTek Innovations LLC, Evanston, IL}
\affiliation
{$^5$Toyota Research Institute}

\begin{abstract}
  The oxygen vacancy formation energy ($\Delta E_{vf}$) governs defect dynamics and is a useful metric to perform materials selection for a variety of applications. However, density functional theory (DFT) calculations of $\Delta E_{vf}$ come at a greater computational cost than the typical bulk calculations available in materials databases due to the involvement of multiple vacancy-containing supercells. As a result, available repositories of direct calculations of $\Delta E_{vf}$ remain relatively scarce, and the development of machine learning models capable of delivering accurate predictions is of interest. In the present, work we address both such points. We first report the results of new high-throughput DFT calculations of oxygen vacancy formation energies of the different unique oxygen sites in over 1000 different oxide materials, which together form the largest dataset of directly computed oxygen vacancy formation energies to date, to our knowledge. We then utilize the resulting dataset of $\sim$2500 $\Delta E_{vf}$ values to train random forest models with different sets of features, examining both novel features introduced in this work and ones previously employed in the literature. We demonstrate the benefits of including features that contain information specific to the vacancy site and account for both cation identity and oxidation state, and achieve a mean absolute error upon prediction of $\sim$0.3 eV/O, which is comparable to the accuracy observed upon comparison of DFT computations of oxygen vacancy formation energy and experimental results. Finally, we demonstrate the predictive power of the developed models in the search for new compounds for solar-thermochemical water-splitting applications, finding over 250 new AA$^{\prime}$BB$^{\prime}$O$_6$ double perovskite candidates.
\end{abstract}

\maketitle

\section{Introduction}\label{Introduction}
Oxygen vacancies are a common defect in metal oxides, and play a key role in several technologies. In solid oxide fuel cells (SOFCs), for example, the vacancy sites are instrumental for bulk oxide ion diffusion, providing pathways for the transport of the ions \cite{Adler2004}\cite{Kilner2014}\cite{Pavone2011}. In solar thermochemical hydrogen (STCH) production cycles, vacancies are created when a metal oxide is reduced at high temperatures generated by concentrated sunlight, and then these vacant sites are filled when the material is re-oxidized by exposing it to steam at a lower (but still elevated) temperature, leaving hydrogen as a product \cite{Nakamura1977}\cite{Wang2012}\cite{McDaniel2013}\cite{McDaniel2014}\cite{Scheffe2013}. The study of oxygen loss can also be of interest in order to avoid detrimental effects, as can be the case in systems such as battery cathodes, where O loss is part of the degradation of the material with consequent loss of function \cite{Gu2013}\cite{Hu2018}.

Numerous efforts have been dedicated to determining the thermodynamic quantities associated with oxygen loss in metal oxides, both experimentally, such as via thermogravimetric measurements, and computationally, utilizing techniques such as density functional theory (DFT) \cite{Panlener1975}\cite{Mizusaki1985}\cite{Mizusaki1989}\cite{Nowotny1998}\cite{Takacs2016}\cite{Grieshammer2013}\cite{xin}\cite{xin2}\cite{emanuela2}\cite{Emery2016} \cite{Ezbiri2017} \cite{Vieten2019} \cite{Gautam2020} \cite{Wexler2021} \cite{Deml2014a}\cite{Deml2015}  \cite{Tezsevin2021}\cite{my_perovskite_paper}. One of the most important thermodynamic properties is the formation energy of neutral oxygen vacancies ($\Delta E_{vf}$), which can be calculated from the DFT energy difference between a compound with and without a vacancy, appropriately accounting for the chemical potential of the removed oxygen. As such, the DFT computation of $\Delta E_{vf}$ has been the subject of a number of studies (see \textit{eg}. \cite{Emery2016} \cite{Vieten2019} \cite{Gautam2020} \cite{Wexler2021} \cite{Deml2014a}\cite{Deml2015}  \cite{Tezsevin2021}\cite{my_perovskite_paper}), with promising results on its accuracy when compared to experimental data \cite{my_perovskite_paper}.

The oxygen vacancy formation energy is a key metric that can be used to guide materials selection for multiple of the previously mentioned applications. In STCH, in particular, new candidate metal oxides can be vetted by screening for materials with $\Delta E_{vf}$ large enough to split water but not so large as to require unreasonable temperatures to create vacancies \cite{Meredig2009}. In order to effectively utilize screening strategies based on $\Delta E_{vf}$ to search for new materials on a large scale, sizeable datasets of defect energies will be required. Compared to defect-free bulk DFT calculations however, which are relatively inexpensive and publicly available for 10$^5$-10$^6$ compounds in databases such as the Open Quantum Materials Database (OQMD) \cite{Saal2013}\cite{Kirklin2015}, the Materials Project \cite{Jain2013}, and Automatic−FLOW for Materials Discovery (AFLOW) \cite{Curtarolo2012}, $\Delta E_{vf}$ calculations are significantly more expensive as they can often require the creation of supercells and the consideration of multiple possible vacancy sites, as well as having lower symmetry. This higher computational cost prohibits the use of direct DFT to produce extremely large datasets (i.e. of dimensions comparable to those of defect-free structure energies) of $\Delta E_{vf}$.

Hence, in order to screen large pools of candidate materials, the development of more computationally efficient ways to generate $\Delta E_{vf}$ predictions is of significant interest. Over the last few years, machine learning models of growing complexity predicting quantities such as the pristine compounds' formation energy have proven effective in drastically reducing DFT computation costs \cite{Ward2017} \cite{Xie2018} \cite{Jha2018a} \cite{Park2020} \cite{Pal2022}. Defect energy prediction, however, is rendered complex by multiple factors, such as the sensitivity to the local bonding environment and the different nature of bonding across compounds and even within the same structure. 

A number of previous studies attempted to identify predictors of $\Delta E_{vf}$ which would allow bypassing DFT defect calculations entirely, relying on properties available from DFT calculations of the defect-free structure, and in some cases even proposing structural and compositional descriptors that do not require any DFT calculation at all. Deml et al. developed two different models of $\Delta E_{vf}$ based on a linear combination of descriptors \cite{Deml2014a} \cite{Deml2015}. The first work \cite{Deml2014a} leveraged a training set of 10 perovskite oxides and utilized the bulk material's formation enthalpy and bandgap as descriptors. The subsequent work \cite{Deml2015} expanded the training set to 45 binary and ternary oxides with a variety of crystal structure types, and expressed $\Delta E_{vf}$ as a function of the formation enthalpy, the average difference in Pauling electronegativity between the oxygen and its nearest neighbor cations, and the energy difference between the oxygen p-band center and the middle of the band gap. Wan et al \cite{Wan2021} then utilized the model predictions on 1750 oxides in Deml et al.'s 2015 work \cite{Deml2015} to identify simple descriptors of $\Delta E_{vf}$ solely based on compositional features that do not require a DFT calculation of the pristine material. They tested several regression methods and identified the best descriptors to be  the difference in Pauling electronegativity between the oxygen and its nearest neighbor cation and the fraction of valence electrons in the material belonging to oxygen. Leveraging instead the dataset of around 300 $\Delta E_{vf}$ calculations of perovskite oxides by Emery et al. \cite{Emery2016}, Liu et al. \cite{Liu2022} also aimed to identify simple features not requiring any DFT calculation, considering elemental properties and proposing a descriptor composed of cation valence, electronegativity, and atomic radii. Recently, Wexler et al. \cite{Wexler2021} introduced a description of $\Delta E_{vf}$ as a linear combination of DFT defect-free stability, band gap, reduction energy of the metal cations neighboring the vacancy, and bond strength between the oxygen and its neighboring metal cations. With such features, they presented two linear models, one trained on a set of 142 oxygen vacancies of perovskite oxides lying within 25 meV/atom of the convex hull, and another one trained on perovskites in any of 6 considered distortions without any stability restriction. Approaching the problem from a more general point of view, Witman et al \cite{Witman2023} selected $\sim$200 different oxides spanning 63 space groups and 14 elements and computed the formation energy of all unique defects in each structure, thereby generating a dataset of over 1000 defect energies. The authors then employed a significantly more complex type of learning than any of the previously mentioned studies, adapting the CGCNN model originally introduced by Xie et al. \cite{Xie2018} to predict any type of defect formation energy by including oxidation state and site-specific information. 

The examples of models of $\Delta E_{vf}$ present in the literature contain significant differences with regards to the training set, the features, and the type of models employed to describe the vacancy formation energy. All three of these factors can significantly affect $\Delta E_{vf}$ predictions in terms of accuracy, generality, interpretability, and training time. 

\textit{Features}: one can broadly divide feature types in two categories: (i) global, i.e. quantities that are single-valued for each defect-free compound, such as the formation energy and the bandgap, and (ii) site-specific, or local, i.e. that depend on the specific O vacancy site under consideration, and are multi-valued for each defect-free compound, such as any feature that depends on the nearest neighbors of the vacancy site.  Models only containing global features are limited in that, since such features don’t allow for the differentiation of single vacancy sites, the model  will give the same prediction for all O vacancy sites in the same material. Furthermore, features related to the metal cations can be further divided into either (i) element-specific, such as the Pauling electronegativity, or (ii) element- and valence-specific such as the ionic radius. Models that heavily rely on non-valence-specific cation features are limited in their ability to differentiate the behaviour of compounds containing the same elements but in different oxidation states. 

\textit{Training Set}: models with smaller training sets containing less structural and compositional variety will be less likely to give accurate predictions when extrapolating to new structures and compositions that are less represented in the training data. Furthermore, training sets containing structures that are significantly off the convex hull or are not the lowest-energy structure at each composition are more likely to contain dynamically unstable pristine structures, which present significant complications when computing the DFT oxygen vacancy formation energy\cite{my_perovskite_paper}.

\textit{ML Model}: linear models, while being fast and intuitive, do not capture more complex relationships between descriptors and target properties, are sensitive to outliers, and can be prone to overfitting. On the other hand, non-linear models are capable of identifying useful feature representation by leveraging complex relationships between variables and can perform better upon extrapolation, but, particularly in the case of neural networks, come at a  greater cost in terms of training time and interpretability, and typically require a larger amount of training data to perform well.

\begin{figure*}[!ht]
\includegraphics[width=14.5cm]{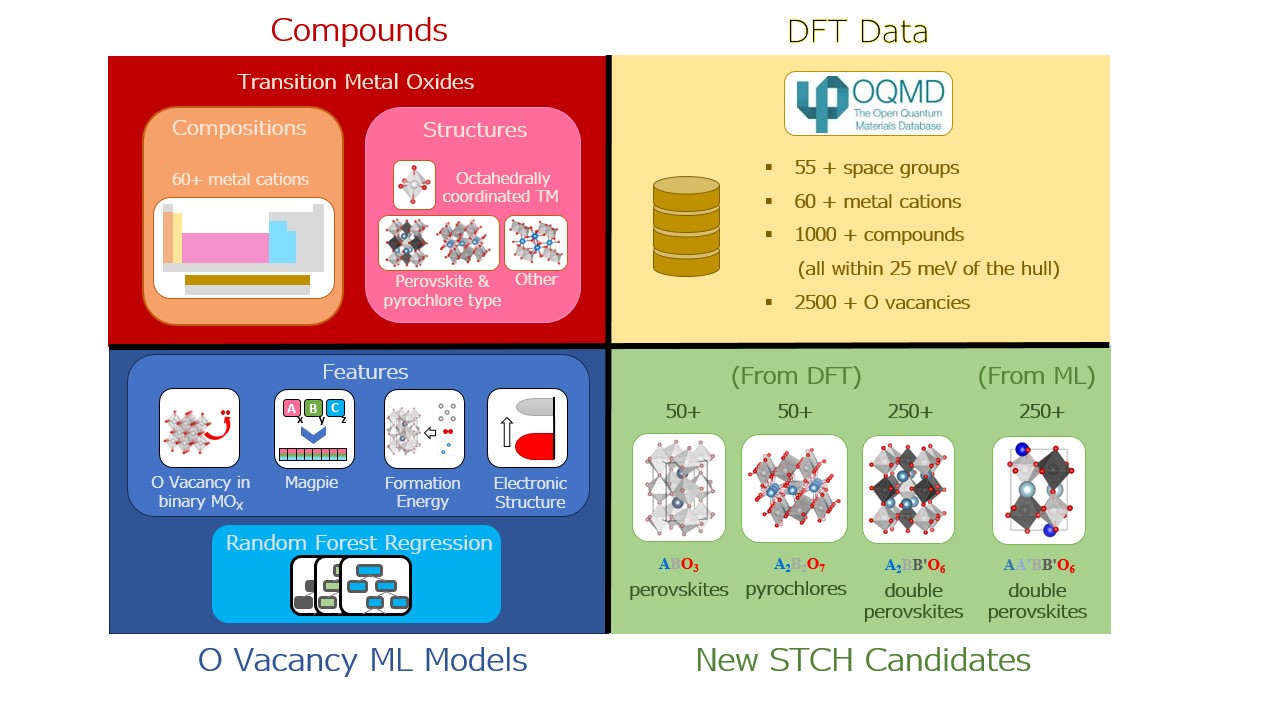}
	\caption{Visual description of the main aspects of the study: (i) high-throughput DFT calculations of oxygen vacancy formation energy (top left and right quadrants), (ii) machine learning model development (bottom left), and (iii) new STCH candidates (bottom right).}
\label{Summary_figure}
\end{figure*}

In this study we tackle all the above discussed aspects of oxygen vacancy formation energy prediction. We first generate a new pool of over 2500 DFT-calculated vacancy formation energies of over 1000 different materials with a variety of structures and compositions, all (in their pristine form) within 25 meV/atom of the convex hull. This constitutes the largest dataset of directly computed DFT vacancy formation energy to date, to our knowledge. A large portion of high-throughput $\Delta E_{vf}$ calculations are conducted on newly predicted perovskite-type and pyrochlore-type compounds found in recent studies \cite{jiangang_double_perov}\cite{jiangang_pyro}, and, among such compounds, we identify 300+ new STCH candidates. We then utilize the newly generated dataset to train machine learning models (random forest, vector, kernel ridge, and linear regression), introducing new descriptors of $\Delta E_{vf}$ and comparing their performance with that of the descriptors previously employed in the literature. The presence of information differentiating the specific vacancy site and accounting for both cation identity and oxidation state emerges as particularly useful. Out of the different feature sets and regressors examined, we find a random forest model using the features introduced in this work to achieve the best performance, obtaining a mean absolute error on testing of $\sim$ 0.3 eV/O, comparable to the accuracy observed upon comparison of DFT computations of oxygen vacancy formation energy and experimental results \cite{my_perovskite_paper}. We subsequently leverage the model's predictions to aid in the search for materials for STCH applications, successfully identifying over 250 new AA$^{\prime}$BB$^{\prime}$O$_6$ double perovskite candidates. Figure \ref{Summary_figure} provides a visual representation of all aspects of the work.

The work is divided as follows. In Section \ref{Data} we present the results of $\Delta E_{vf}$ calculations from high throughput DFT studies of several different structure types. In Section \ref{Features} we introduce and discuss several features to be used for $\Delta E_{vf}$ prediction. In Section \ref{Models} we then utilize our large DFT dataset to train different machine learning models, comparing the performance of different regressors and different sets of features. Finally, in section \ref{Predictions} we utilize the model to predict new AA$^{\prime}$BB$^{\prime}$O$_6$ double perovskite STCH materials.

\section{Methodology}\label{Methodology}
\paragraph{DFT Calculations}
All DFT calculations in this work have been conducted using the Vienna ab-initio simulation package (VASP)\cite{Kresse1996}\cite{Kresse1996a}, with projector augmented wave (PAW) potentials \cite{Kresse1999} and the Perdew-Burke-Ernzehof (PBE) \cite{Perdew1996} generalized gradient approximation (GGA) for the exchange-correlation functional. The calculations were conducted within the framework of OQMD,  details on the settings employed can be found in Refs \cite{Saal2013}\cite{Kirklin2015}. For compounds containing 3d transition metals or actinides, the over-delocalization of electrons due to the residual self interaction present in exchange-correlation functionals \cite{DFTUreview}\cite{Zhou2004}\cite{Wang2006}\cite{Lutfalla2011}\cite{Chevrier2010} was treated through the addition of a Hubbard-like potential to the energy functional \cite{Dudarev1998}, and a spin polarization was applied with a ferromagnetic configuration, initializing magnetic moments to 5$\mu_B$ (transition metals) and 7 $\mu_B$ (actinides). The oxidation state of each cation was determined using bond valence parameters as implemented in pymatgen \cite{Ong2013}, and, for a small number of compounds, that contained a rare earth element whose oxidation state differed from that of the pseudopotential used in the OQMD framework, the vacancy formation energy was not calculated.
\paragraph{Stability and Oxygen Vacancy Formation Energy}
The compounds' 0K stability ($\Delta E_{stab}$) was calculated by taking the energetic difference between the OQMD convex hull energy (calculated excluding the composition of interest) at the composition of interest and the formation energy of the compound($\Delta E_{f}$). The details of stability calculations are reported in other works \cite{my_perovskite_paper}\cite{Emery2016}\cite{jiangang_double_perov}\cite{jiangang_pyro}\cite{Griesemer2021}.
\\*The oxygen-vacancy formation energy ($\Delta E_{vf}$) was then determined by removing an oxygen atom from a supercell of the relevant bulk structure, and computed as follows:
\begin{equation}
	\Delta E_{vf} = E(A_{na}B_{nb}C_{nc}O_{no-1})  + E_O - nE(A_aB_bC_cO_o)
\label{evf}
\end{equation}
where  $E(A_aB_bC_cO_o)$ indicates the DFT energy of a bulk metal oxide unit cell, $E(A_{na}B_{nb}C_{nc}O_{no-1})$ that of a supercell n times the size of the bulk unit cell and containing one oxygen vacancy, $E_O$ is the DFT reference energy of oxygen (which corresponds to the corrected DFT elemental energy, the OQMD correction resulting in a value of  $E_O$=-4.523 eV, as discussed in more detail in Refs. \cite{Saal2013}\cite{Kirklin2015}\cite{my_perovskite_paper}), and the units of $\Delta E_{vf}$ are $eV/O$.
In the interest of selecting structures more likely to be synthesized and avoiding structures with a significant degree of dynamic instability \cite{my_perovskite_paper}, for each composition, $\Delta E_{vf}$ calculations were conducted with the lowest-energy structure at that composition. For each structure, one vacancy-containing supercell with a minimum of 15 atoms (see Fig S1 for cell size convergence) per unique oxygen site was created. Convergence issues were encountered for a small number of the defect calculations, which are therefore not included in the dataset. Since we utilize site-specific vacancy formation energies (rather than, eg. minima or averages for each compound) to train and test the machine learning models, the above mentioned convergence issues of a small number of vacancy calculations does not impact the accuracy of the results presented in the work. 
In part of our analysis, we consider two different contributions to the oxygen vacancy formation energy, one from the removal of the O atom alone (but otherwise not relaxing the remaining atoms), and the other from the relaxation of other ions upon the removal of the O atom. $\Delta E_{vf}$ is then given by the sum of these two terms \cite{my_perovskite_paper}:
\begin{equation}
	\Delta E_{vf} = \Delta E_{vf}^{UN} + \Delta E_{vf}^{R}
\label{evf_rel_un}
\end{equation}

$\Delta E_{vf}^{UN}$ indicates the "unrelaxed" contribution, which corresponds to the energetic cost of creating an O vacancy before allowing for any relaxation of the-vacancy containing cell:
\begin{equation}
	\Delta E_{vf}^{UN} = E(A_{na}B_{nb}C_{nc}O_{no-1})^{UN} +E_O - nE(A_aB_bC_cO_o)
\label{evf_un}
\end{equation}
where $E(A_{na}B_{nb}C_{nc}O_{no-1})^{UN}$ indicates the energy of the defect containing cell with cell and atomic positions fixed to what they were in the relaxed pristine cell, and $E(A_aB_bC_cO_o)$ is the energy of the relaxed pristine cell

$\Delta E_{vf}^{R}$ indicates the "relaxation" contribution, which corresponds to the change in energy of the vacancy containing cell when allowing for relaxation of the cell and the ionic positions:
\begin{equation}
	\Delta E_{vf}^{R} = E(A_{na}B_{nb}C_{nc}O_{no-1})^{R} - E(A_{na}B_{nb}C_{nc}O_{no-1})^{UN}
\label{evf_r}
\end{equation}
where $E(A_{na}B_{nb}C_{nc}O_{no-1})^{R}$ indicates the energy of the defect containing cell after allowing for cell and atomic position relaxation. Observations of large $\Delta E_{vf}^{R} $ are likely an indication of significant structural changes.
\paragraph{Machine Learning}
We developed models of vacancy formation energy utilizing four different regression techniques: random forest regression (RFR) \cite{Breiman2001}, support vector regression (SVR) \cite{Zhang2020}, kernel ridge regression (KRR) \cite{Vu2015} and linear regression (LR). In all cases, the models were developed utilizing the Scikit-learn Python library, and optimizing the hyperparameters through a grid search (200 estimators for RFR, and a radial basis function kernel for SVR and KRR).
A part of the features we computed rely on the determination of nearest neighbors and oxidation states. Nearest neighbors were determined using the CrystallNN algorithm implemented in pymatgen \cite{Ong2013}, and oxidation states were determined utilizing a bond valence method, again as implemented in pymatgen \cite{Ong2013}.

\section{High Throughput DFT Oxygen Vacancy Formation Energies}\label{Data}
The present section is dedicated to presenting the results of DFT calculations of oxygen vacancy formation energy of over 1000 different compounds. The dataset formed by such calculations is then used as a training set for the machine learning models presented in the next section. The results of the calculations themselves, however, being largely novel and unpublished, and having led to the identification of several new STCH candidate materials, merit a separate focused discussion.

The dataset can be divided into two main parts: one containing the results of $\Delta E_{vf}$ calculations of compounds within 25 meV of the OQMD convex hull with specific structure types (namely: perovskite, pyrochlore and double perovskite) identified through our previous targeted high-throughput DFT studies, and a second one containing the results of $\Delta E_{vf}$ calculations of compounds with multiple different structure types (all with octahedral coordination of transition metal cations) already present on the OQMD convex hull. For each composition, the oxygen vacancy formation energy is calculated utilizing the lowest energy structure present on the OQMD at that composition. Furthermore, no composition for which the lowest energy structure is more than 25 meV/atom off the hull is present in the dataset. We will first examine in detail the results from the high-throughput studies individually, then give a collective vision of all the calculations.

\subsection{Perovskites}
In a 2016 study of perovskite oxides, Emery et al. \cite{Emery2016} computed the DFT stability of all decorations of the ABO$_3$ perovskite structure (with A and B being metal cations), considering the four most common perovskite distortions. The authors identified over 250 stable (within 25 meV/atom of the convex hull) compounds, with the rare earths and alkaline earth metals predominantly occupying the A site, and transition metals the B site. In a recent work \cite{my_perovskite_paper} we updated the dataset of stable ABO$_3$ compounds following the addition of numerous new compounds to the OQMD (now containing over 1 million entries \cite{oqmd_million}), and computed the vacancy formation energy of the lowest energy structure at each stable ABO$_3$ composition, highlighting a strong dependence of $\Delta E_{vf}$ on the B site cation. 

We illustrate in Figure \ref{evf_peorv_pyro} the results of the vacancy formation energy calculations of perovskite oxides, plotted against $\Delta E_{stab}$ to highlight the previously mentioned 25 meV/atom stability cutoff. Due to the asymmetry in the oxygen octahedra surrounding each B cation present for distorted phases (such as the orthorhombic phase, which is the lowest energy structure in most cases), multiple non-equivalent O vacancy sites can be identified in each structure. The difference between the sites is related to the length and angle of the bonds between the oxygen and the metal cation, and the energetic difference between the sites rarely exceeds 0.2 eV/O, so we only represent the lowest of the two values for each compound. We remark the significant dependence of $\Delta E_{vf}$ on the element and oxidation state of the B site cation, which is indicated with different markers. As discussed in \cite{my_perovskite_paper}, the B cation is also the one where the majority of the charge localizes upon formation of the O vacancy, and the relative trends in $\Delta E_{vf}$ can be associated with an 'ease of reduction' of the different cation (with, for example, Mn$^{4+}$ being easier to reduce than Cr$^{3+}$). We also note that, while in the majority of cases the A sites are occupied by rare-earth or alkali metal atoms, a small number of compounds with eg. Bi$^{3+}$ or Te$^{3+}$ on the A site are also present, with $\Delta E_{vf}$ being noticeably lower for such compounds when compared to ones having the B cation and a rare-earth A cation. Such cases are also the ones in which a larger portion of the charge localizes on the A cation. 

\subsection{Pyrochlores}
In a recent study, He et al. \cite{jiangang_pyro} computed DFT the stability of all possible A$_2$B$_2$O$_7$ compositions (with A and B being metal cations) in the pyrochlore structure, as well as the 2 competing structures most commonly found in the Inorganic Crystal Structure Database (ICSD) \cite{Bergerhoff1983}\cite{Belsky2002}. More than 300 stable (25 meV/atom of the hull) compounds were identified, with the A site being predominantly occupied by rare earth elements and the B site by transition metals. Upon computing the vacancy formation energy, 60 such compounds were found to lie within the 2-5 eV/O window of interest for STCH applications \cite{my_perovskite_paper}, a list of such compounds is provided in Table S1.

\begin{figure*}[!ht]
\includegraphics[width=16.5cm]{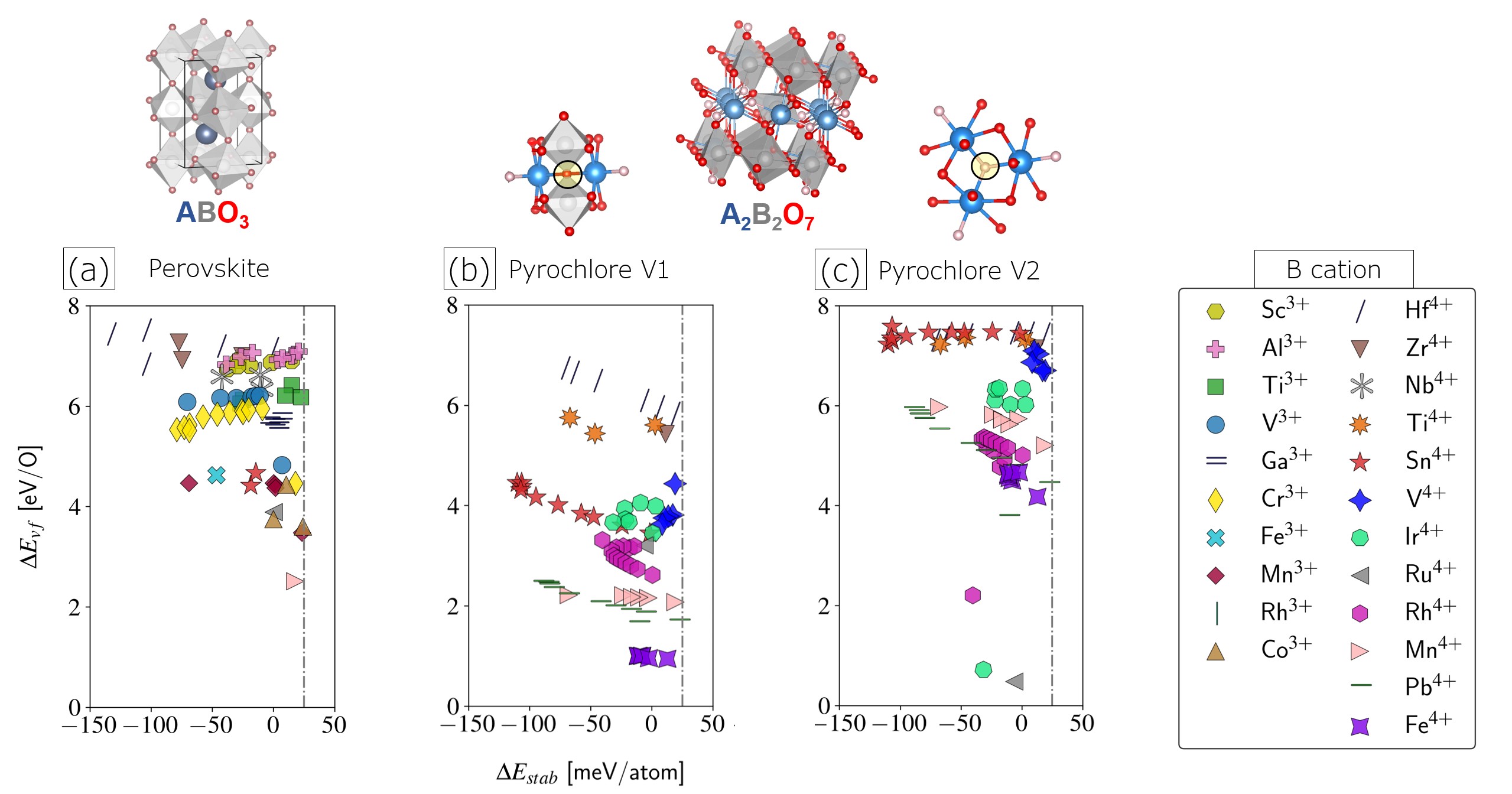}
	\caption{Oxygen vacancy formation energy of (a) ABO$_3$ perovskite oxides, and (b) and (c) A$_2$B$_2$O$_7$ pyrochlore oxides. In (a) $\Delta E_{vf}$ of the lowest energy vacancy is reported, as all vacancy sites have the same A and B cation nearest neighbors. In (b) $\Delta E_{vf}$ of sites with A and B cations (where B cations are the closest) as nearest neighbors is reported, and in (c) $\Delta E_{vf}$ of sites with only A cation nearest neighbors is reported. In all cases different markers and colors indicate the species and oxidation state of the B cation of each given compound. The vertical line indicates the 25 meV/atom stability ($\Delta E_{stab}$) cutoff applied for compound selection.}
\label{evf_peorv_pyro}
\end{figure*}

In Figure \ref{evf_peorv_pyro} b and c we present the results of oxygen vacancy formation energy calculations of stable pyrochlore compounds. Like the perovskites examined above, pyrochlore oxides have multiple non-equivalent vacancy sites. In this case, however, one oxygen type forms octahedra around B cations, and another one forms tetrahedra around A cations. The difference between the sites, therefore, is not just in the lengths and angles of the bonds between the oxygen and its nearest neighbor cations, but also in the species of the neighboring cations. For one of the sites, the nearest neighbors (NN) to the O atom are B cations (as well as, at larger distances, A cations), while for the other the nearest neighbors are A cations. Both vacancy sites are represented in Figure \ref{evf_peorv_pyro}: the site with B cations as NN (also referred to as "V1") in Figure \ref{evf_peorv_pyro} b, and the site with A cations as NN (also referred to as "V2") in NN Figure \ref{evf_peorv_pyro} c, with different markers indicating the B site cations in both cases. A significant difference can be observed between the two sites, with the vacancy formation energy of sites surrounded only by rare earth cations being significantly higher than that of the sites with transition metal neighbors, which in turn has similar values to perovskites with the same B transition metal cation neighbors.

\subsection{Double Perovskites}
In a wide ranging study comprising over 35,000 calculations, He et al. \cite{jiangang_double_perov} computed the stability of B site ordered double perovskites in the rock salt-type double perovskite structure. Such structure is the most commonly observed one for A$_2$BB$^{\prime}$O$_6$ double perovskites and is characterized by an alternation between the B and B' cation on both rows and columns \cite{Anderson1993}\cite{King2010}\cite{Howard2003} . To determine the lowest energy phase, 10 different distortions of A$_2$BB$^{\prime}$O$_6$ perovskites were considered (P21/c, C2/m, P1, P-1,  I4/m, I4/mmm, R3, R-3, R-3m, Fm-3m), with A=Ca, Sr, Ba and La (and, for the most common P21/c structure, A=Zn, Cd, Hg and Pb), and 50 metal elements on the B sites, totaling around 10,000 compositions. The 3 most common  A$_2$BB$^{\prime}$O$_6$ competing phases on the ICSD were also calculated. Using this strategy, more than 500 stable (within 25 meV/atom of the convex hull) double perovskites were identified. For each of those compounds, we computed the vacancy formation energy of all non-unique oxygen sites, finding over 250 new STCH candidates, a list of which is provided in Table S2.

Once again, a strong influence of the identity of the cations neighboring the vacancy on $\Delta E_{vf}$ can be observed. Due to the nature of the rocksalt-type B site ordering of the double perovskite structure, each unique O is bonded to both B site cation species (B and B'). Examining cases in which both B and B' are among the species highlighted in Figure \ref{evf_peorv_pyro}, we observe similar trends to those identified in simple perovskites and pyrochlores. For example, among La$_2$AlB'O$_6$ double perovskites, having B'=V$^{3+}$, Cr$^{3+}$, and Fe$^{3+}$ leads to a progressively lower vacancy formation energy (with values of, respectively, 6.2 eV/O, 5.8 eV/O and 4.4 eV/O). Similarly, Mn$^{4+}$ is associated with a lower $\Delta E_{vf}$ than Sn$^{4+}$ when occupying the B' site in Ca$_2$ZrB'O$_6$ (2.8 eV/O vs 4.7 eV/O). While the above examples highlight the dominant role in determining $\Delta E_{vf}$ to be played by the easiest to reduce cation, significant differences in $\Delta E_{vf}$ can also be identified when the identity of the other cations is changed. Once again, the trends identified in simple perovskites and pyrochlores appear to hold, with, for example, B=Hf$^{4+}$, Zr$^{4+}$, and Ti$^{4+}$ leading to a progressively lower $\Delta E_{vf}$ in Ca$_2$BMnO$_6$ (with values of, respectively 3.0 eV/O, 2.8 eV/O and 2.3 eV/O)

\subsection{Other Metal Oxides}
In addition to the data produced through the high-throughput studies presented above, we also screened the OQMD for metal oxide compounds satisfying the following criteria: (i) the compound lies on the convex hull, (ii) the compound contains transition metal cations, (iii) oxygen is the only anion species, and (iv) the transition metal cations have octahedral coordination to oxygen anions. Of these compounds, about half were reported on the ICSD, and the other half consists of hypothetical compounds calculated to be DFT stable. The hypothetical compounds were discovered exploiting several different strategies: solution of experimental diffraction data using database searching \cite{Griesemer2021}, machine learning prediction leveraging an improved version of CGCNN \cite{Park2020}, substitution of chemically similar elements into already-known compounds \cite{Glawe2016}\cite{Hautier2011}, and inclusion of compounds discovered in the work of Wang et al. \cite{Wang2021}.
For all compounds, we computed the oxygen vacancy formation energy of different unique vacancy sites.

\subsection{Entire Dataset}
\begin{figure}[!ht]
\includegraphics[width=8.3cm]{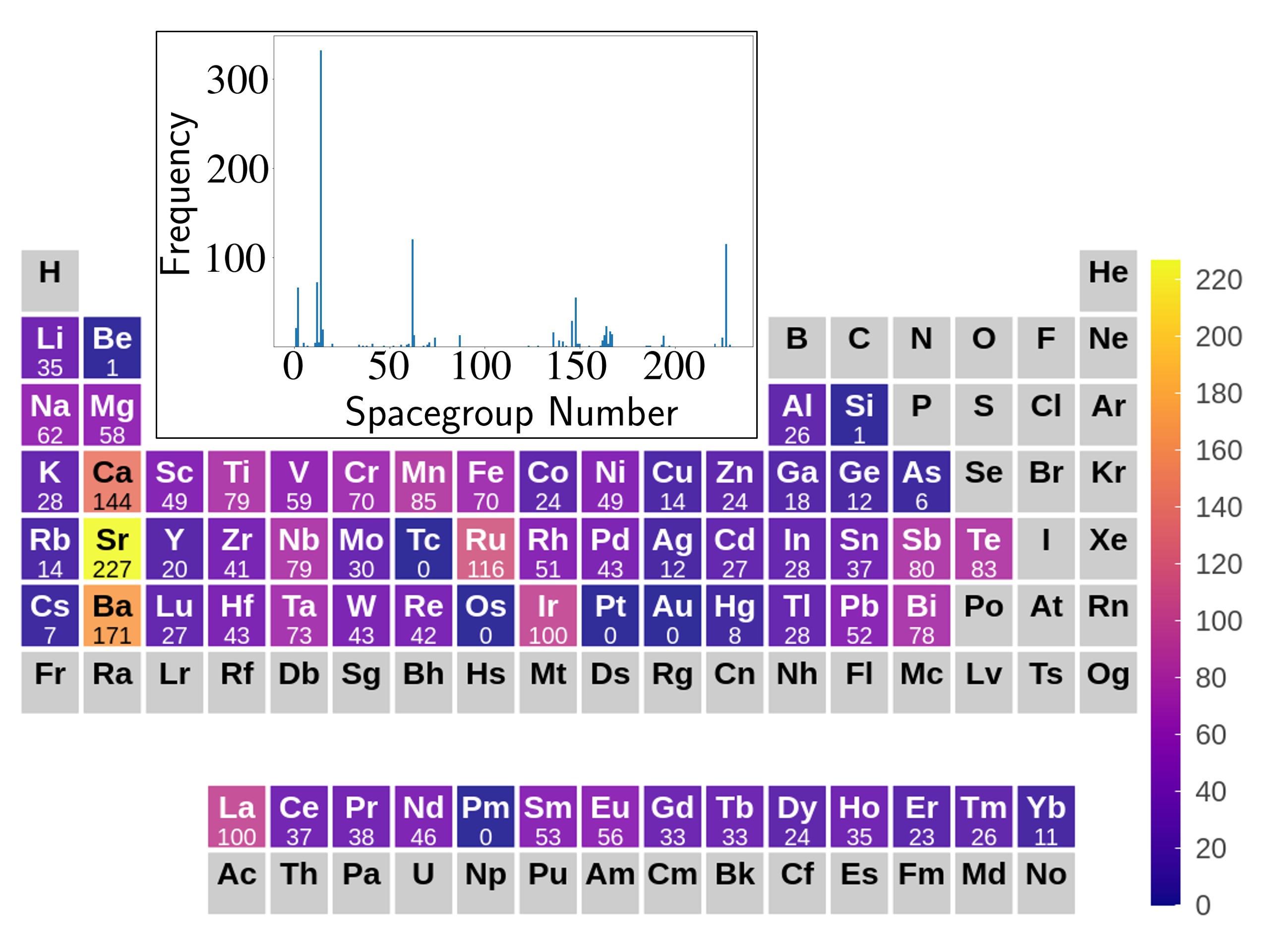}
	\caption{Distribution of the compounds in the entire dataset over elements (lighter shades indicating more compounds containing that element) and spacegroups (where each spacegroup is indicated by its number index). 55 different spacegroups are represented in the dataset, the most frequent ones being the ones that perovskite, pyrochlore and double perovskite structures belong to. Alkali, alkali earth, rare earth, transition metal, post transition metal and metalloid cations are present in the dataset, with Ca, Ba and Sr having the largest frequency as they often serve as A site cations in the A$_2$BB$^{\prime}$O$_6$ double perovskites compounds.}
\label{dataset}
\end{figure}

Combining the data from all the above-described calculations, our entire dataset contains 2677 different oxygen vacancy formation energy calculations of 1157 different metal oxides.
Figure \ref{dataset} shows the distribution of the entire dataset over spacegroups and elements. Structures belonging to 55 different spacegroups appear in the dataset, with the highest frequency spacegroups being P21/c (to which most of the A$_2$BB$^{\prime}$O$_6$ double perovskites belong to), Pnma (to which most ABO$_3$ perovskites belong to), and Fd-3m (to which A$_2$B$_2$O$_7$ pyrochlores belong to). 66 different cation elements are considered, with 90\% appearing at least 5 times in the dataset, and 75\% appearing at least 20 times. 

\begin{figure}[!ht]
\includegraphics[width=8.3cm]{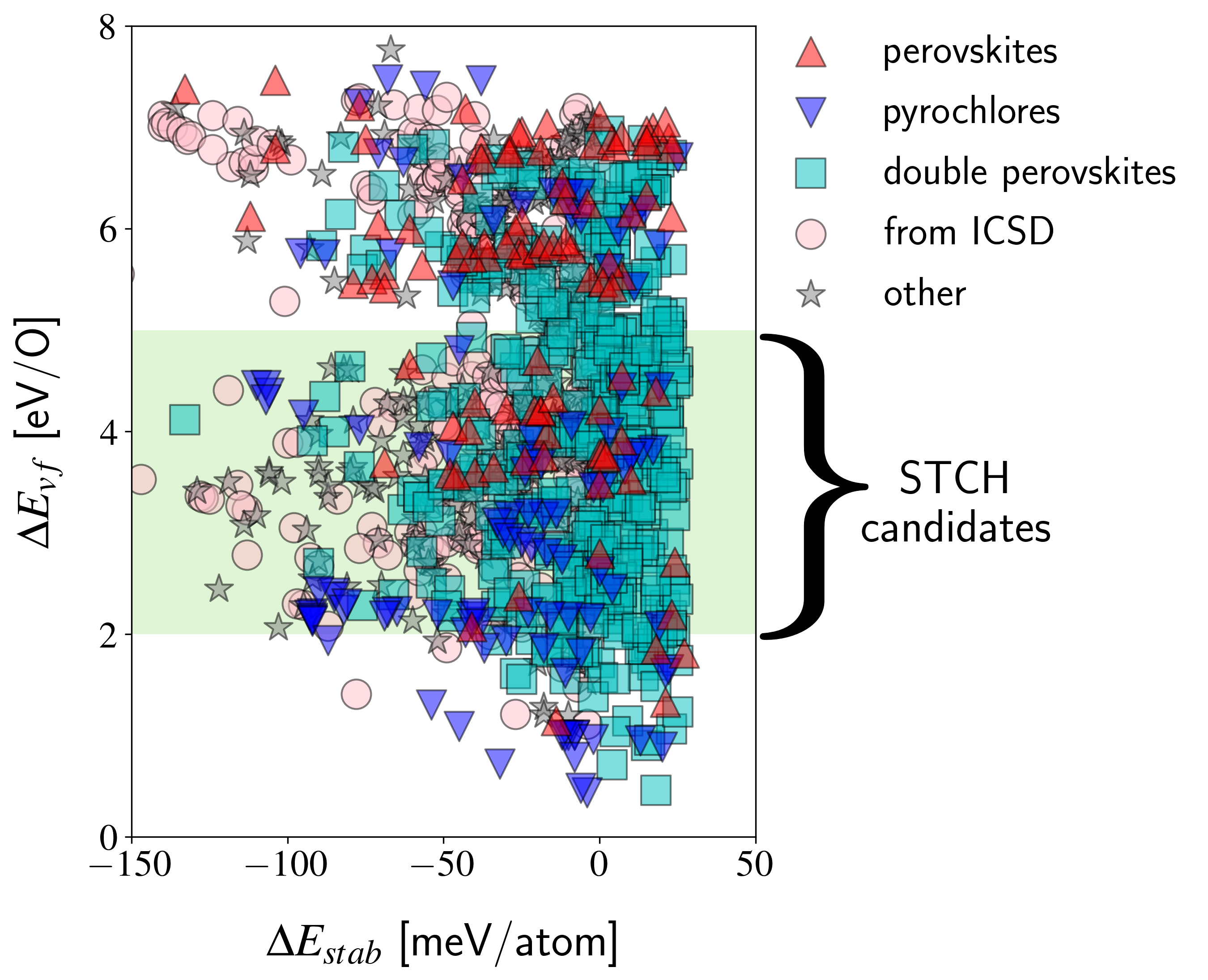}
	\caption{Oxygen vacancy formation energy of the compounds calculated within the high throughput study of perovskites (red triangles), pyrochlores (blue triangles) and double perovskites (cyan squares), as well as compounds on the OQMD convex hull with octahedrally coordinated transition metal cations, both from the ICSD (pink circles) and from other studies (black stars). The window of interest of oxygen vacancy formation energy for STCH applications is highlighted in green.}
\label{STCH_candidates_all}
\end{figure}

Figure \ref{STCH_candidates_all} displays all the compounds in the dataset as individual datapoints (the lowest value of $\Delta E_{vf}$ for each compound is shown), with different markers indicating the compounds calculated within the high throughput study of perovskites (red triangles), pyrochlores (blue triangles) and double perovskites (cyan squares), as well as compounds on the OQMD convex hull with octahedrally coordinated transition metal cations, both from the ICSD (pink circles) and from other studies (black stars). As previously mentioned, the high throughput studies have also led to the identification of hundreds of new STCH candidates, highlighted in the figure through a green banner. Lists of the new pyrochlore and double perovskite candidates are provided in the SI, and a list of ABO$_3$ perovskite candidates was included in our recent publication \cite{my_perovskite_paper}. The rest of the dataset will be publicly available following the publication of our other recent works \cite{jiangang_double_perov} \cite{jiangang_pyro}.

\section{Machine Learning Models}\label{ML}
In developing a model for predicting the oxygen vacancy formation energy we have multiple goals. We firstly aim to identify simple-to-extract features that only rely on compositional and generic structural information and can be combined with the widely employed Magpie feature set introduced by Ward et al. \cite{Ward2016} to specifically target oxygen vacancy formation energy prediction. We then also want to leverage the wealth of information generated through DFT calculations of the pristine structure of interest to compose a feature set containing only a small number of descriptors with intuitive relation to oxygen vacancy formation energy. In both cases, we then want to use the features in conjunction with prediction algorithms capable of identifying relationships between features and target property that go beyond linearity while still being fast and interpretable, and of achieving reliable predictive accuracy starting with even just a few hundred training points. We introduce the selected features in Section \ref{Features}, where we also describe other features previously employed in the literature which we will use for comparison. In Section \ref{Models} we then train random forest models on the data introduced in Section \ref{Data} and compare the performance obtained using various combinations of features, including previous feature sets employed in the literature. We find the models using the feature sets introduced in this work to be the best performing ones out of the ones examined, achieving a mean absolute error as low as $\sim$0.3 eV/O. Finally, in Section \ref{Predictions} we employ such models to predict the oxygen vacancy formation energy of newly calculated, DFT stable (within 25 meV/atom of the convex hull), A and B site ordered AA$^{\prime}$BB$^{\prime}$O$_6$ double perovskites and select the most promising ones for STCH applications.  

\subsection{Features}\label{Features}
The present section is dedicated to the description of various features used for $\Delta E_{vf}$ predictions. First, the choices behind the two feature sets selected in this work (referred to as Feature set I and Feature set II) are detailed. Then, other features used in previous works are listed and briefly described, explicitly mentioning which ones can be classified as global features, and which as site-specific. The nomenclature introduced in this section to refer to each feature and feature set is utilized for the rest of the paper.

Our first step is to introduce features capable of accounting for the strong influence of the atomic species and oxidation state of the cations in each metal oxide on the oxygen vacancy formation energy, which we have highlighted in Section \ref{Data}. In order to differentiate each cation in our dataset while capturing the energetic change involved in the breaking of the bonds and the redistribution of charge involved in the creation of an oxygen vacancy, we compute the unrelaxed oxygen vacancy formation energy (see Methods and Ref \cite{my_perovskite_paper}) of a binary oxide containing each such cation. We refer to this quantity as $\Delta E_{vf}^{MO_x}$, where M indicates the cation species, the oxidation state is $2x+$, and the binary oxides are selected by looking for the lowest energy MO$_x$ compounds in the OQMD. The values of $\Delta E_{vf}^{MO_x}$ for the cations appearing in this work's dataset are displayed in Figure S2. Trends in accordance with what we observed for perovskites and pyrochlore oxides can be identified: (i) rare-earth and alkaline earth metal species having larger $\Delta E_{vf}^{MO_x}$ than most transition and post-transition metals, (ii) $\Delta E_{vf}^{MO_x}$ decreasing with increasing period among the same transition metal series, and (iii) $\Delta E_{vf}^{MO_x}$ being larger for smaller oxidation states for the same metal.

We then construct two feature sets that leverage the information contained in the descriptor detailed above.

Feature set I contains two types of features: 
\begin{itemize}
    \item Magpie:
    \\* The Magpie feature set introduced by Ward et al., which contains stoichiometric attributes (such as the number of elements in the compound), elemental properties statistics (such as  the average atomic number of the elements in the compound), electronic structure attributes (such as the average number of d electrons across elements in the compound), and ionic compound attributes (such as the "ionic character" of the compound) \cite{Ward2016}. These are all global features.
    \item stats($\Delta E_{vf}^{MO_x}$):
\\*Statistics over features constructed from the unrelaxed vacancy formation energy of binary oxides. Two particular quantities are selected. (i) min($\Delta E_{vf}^{MO_x}$): the lowest value of $\Delta E_{vf}^{MO_x}$ among all $M^{2x+}$ cations in the structure, which is a global feature. Considering, for example, the pyrochlore Lu$_2$Mn$_2$O$7$, where Mn$^{4+}$ is the easiest to reduce cation, min($\Delta E_{vf}^{MO_x}$) would correspond to $\Delta E_{vf}^{MnO_2}$. (ii) $<\Delta E_{vf}^{MOx,NN}>$: the weighted average of $\Delta E_{vf}^{MO_x}$ over all $M^{2x+}$ cations neighboring the vacancy, which is a site-specific feature. The weights are defined to sum to 1, and to be larger for cations with smaller $\Delta E_{vf}^{MO_x}$, so as to reflect the dominant role in determining $\Delta E_{vf}$ that cations with smaller $\Delta E_{vf}^{MO_x}$ have, as observed in Section \ref{Data}. Specifically, for each cation $i$ neighbouring the vacancy having $\Delta E_{vf}^{MO_{x,i}}$, its weight is defined as $w_i = \frac{u_i}{\sum_j u_j}$, with $u_i = \frac{1}{1+(\Delta E_{vf}^{MO_{x,i}}-min(\Delta E_{vf}^{MO_{x,NN}}))} $ , where $min(\Delta E_{vf}^{MO_{x,NN}})$ is the lowest $\Delta E_{vf}^{MO_x}$ among the cations neighboring the vacancy. Considering again the example of pyrochlore Lu$_2$Mn$_2$O$7$, which has two types of vacancy sites (V1 where neighbors are 2 B and 2 A cations, and V2 where neighbors are 4 A cations, as highlighted in Section \ref{Data}), for V1 $<\Delta E_{vf}^{MOx,NN}>$ would correspond to $2w_{MnO_2}\Delta E_{vf}^{MnO_2}+2w_{Lu_2O_3}\Delta E_{vf}^{Lu_2O_3}$ with $w_{Lu_2O_3}= \frac{1}{1+(\Delta E_{vf}^{Lu_2O_3}-\Delta E_{vf}^{MnO_2})} \frac{1}{2u_{MnO_2}+2u_{Lu_2O_3}} \sim 0.1$ and $w_{MnO_2}= 1 \frac{1}{2u_{MnO_2}+2u_{Lu_2O_3}} \sim 0.4$, and for V2 $<\Delta E_{vf}^{MOx,NN}>$ would correspond to $\Delta E_{vf}^{Lu_2O_3}$ . The selection of these quantities is aimed at capturing the influence that the nature of the cations neighboring the vacancy has on the oxygen vacancy formation energy, and reflecting the dominant role of cations with smaller $\Delta E_{vf}^{MO_x}$.
\end{itemize}
A major advantage of only requiring compositional and generic structural information is that this feature set can in principle be utilized without the need for any previous DFT calculation of the bulk structure, and only necessitates a structural prototype decorated with all the elemental combinations of interest (which is often the starting point of high-throughput studies). We note that a significant caveat of the above consideration is that, upon relaxation, atomic positions can at times change enough for the oxidation state of the cations determined (through the bond valence method we utilized, see Methods) from the relaxed and unrelaxed structure to differ. Given that cation oxidation states play a major role in $\Delta E_{vf}^{MOx}$-related features, this can significantly affect predictions. To assess the extent of the impact of this effect, we determine the oxidation states of all cations in all compounds in our dataset before and after relaxation, finding that in 90\% of the cases they remain the same. We therefore conclude that, while not an infallible strategy, Feature set I can be used to coarsely screen $\Delta E_{vf}$ of compounds prior to running DFT calculations. For example, within a search for new compounds with a specific structural prototype(s) and $\Delta E_{vf}$ in a target range, we can use an ML model with Feature set I to produce $\Delta E_{vf}$ predictions on the unrelaxed structures, and then reduce the number of DFT calculations of the pristine structures by only computing structures with a predicted $\Delta E_{vf}$ within a range of interest (ideally larger than the final target range so as to reduce possible false negatives). We will give an example of this strategy in Section \ref{Predictions}. We note that another caveat of the use of the above strategy is the lack of transferability to datasets containing structures presenting complexities such as cation disorder, where the oxidation states of the cations are likely to have greater diversity within the structure and the effects of relaxation to play a larger role.

Feature set II, which excludes Magpie features and is thus much smaller, is composed of the following features:
\begin{itemize}
    \item stats($\Delta E_{vf}^{MO_x}$):
    \\* The same two quantities described in the previous paragraph, with the only difference that the weights used to calculate are now also inversely proportional to the bond length between O and the relevant neighboring cation in order to capture more information on the local environment surrounding the vacancy. These two features are one global and one site-specific.
    \item $\Delta E_{f}$:
    \\* The formation energy of the compound, aiming to capture energy change related to breaking bonds across the whole compound. This is a global feature.
    \item ($E_{Op}^{site}+E_g$):
\\* The energy difference between the center of mass of the p-band of the oxygen atom which becomes vacant and the first unoccupied state in the band structure (as computed using the pristine structure before introducing the vacancy), aiming to capture the energetic change associated with the change of state of the electrons previously involved in the bond with the O atom becoming vacant (assuming a similar band structure after defect creation). This is a site-specific feature.
    \end{itemize}
Having a feature set containing only a small number of features presents the advantage of being versatile, interpretable, and less prone to overfitting risk.

In order to compare the predictive value of our features against features used in other studies, we compute the following features from other studies: 
\begin{itemize}
    \item $V_r^{NN}$:
\\* The maximum reduction potential of the cations neighboring the vacancy. The reduction energy $E_r$ of a metal cation M from M$^{m+}$ to M$^{n+}$ is calculated by taking the energetic difference between the energy of the ground state polymorphs of the binary oxides MO$_{m/2}$ and  MO$_{n/2}$, and accounting for the O energy. The reduction potential is then calculated as $-E_r(M^{m+}\rightarrow M^{n+})/(m-n)F$, where F is Faraday's constant. This is a site-specific feature. This feature was introduced in the work by Wexler et al. \cite{Wexler2021}, where more details can be found. In general, in this work we considered n=m-1, taking the ground state polymorphs from the OQMD. For the minority of cases where the m-1 binary oxide was not on the OQMD convex hull, multiple options were considered: (i) the lowest energy MO$_{(m-1)/2}$ polymorph was utilized regardless of its instability (ii) $V_r$ was quantified with $\Delta E_{f} (MO_{m/2}) / m F$ (corresponding to n=0), and (iii) the next stable binary oxide corresponding to a smaller oxidation state of M (eg. MO$_{(m-2)/2}$) was utilized in place of MO$_{(m-1)/2}$ (corresponding to, eg. n=m-2). The different options were then compared assessing their performance by reproducing the figures in Section \ref{Models} (where option ii is represented), finding the rankings to remain unchanged in all cases. 
    \item $E_{g}$:
\\* The band gap of the bulk compound of interest. This is a global feature. This feature was utilized in the work by Wexler et al. \cite{Wexler2021}, where more details can be found.
    \item $\Sigma E_b^{NN}$:
\\* The sum of the crystal bond dissociation energy of the cations neighboring the vacancy. For each M with oxidation state m, the crystal bond dissociation energy is calculated by taking the ratio between the cohesive energy of the ground state polymorph of the binary oxide MO$_{m/2}$ and the number of M-O bonds present in that binary oxide. In turn, the cohesive energy of a binary oxide MO$_{m/2}$ is calculated by subtracting the cohesive energy of the metal and the bond dissociation energy of O$_2$ from the formation energy of the binary oxide. This is a site-specific feature. This feature was introduced in the work by Wexler et al. \cite{Wexler2021}, where more details can be found. Once again,  we note that in the present work the ground state polymorphs are taken from the OQMD.
    \item $\Delta E_{stab}$:
\\* The stability of the bulk compound of interest. This is a global feature. This feature was utilized in the work by Wexler et al. \cite{Wexler2021}, where more details can be found. 
    \item ($E_{Op}+0.75E_g$):
\\* The sum of the bandgap and the difference in energy between the Fermi energy and the center of mass of the oxygen p-band of a compound . This is a global feature. This feature was introduced in the work by Deml et al. \cite{Deml2015}, where more details can be found (the factor of 0.75 was determined through a fitting process).
    \item $<\Delta\chi^{NN}>$:
\\* The average difference in Pauling electronegativity between oxygen and its first nearest neighbor. This is a global feature. This feature was introduced in the work by Deml et al. \cite{Deml2015}, where more details can be found, and was also employed in the work by Wan et al.  \cite{Wan2021}
    \item $\Delta\chi^{1stNN}$:
\\* The site specific difference in Pauling electronegativity between oxygen and its closest cation. This is a site-specific feature. When evaluating the performance of the features utilized in the works by Deml et al. \cite{Deml2015} and Wan et al.  \cite{Wan2021} in the next section, this feature, rather than the previous one is utilized. The reason behind this choice is that, while the original works only predict one $\Delta E_{vf}$ value for each compound, in this work we separately predict the values of different vacancy sites, and we therefore consider the substitution of the average electronegativity difference with the site specific one to provide a fairer evaluation in this case. We also note that this substitution improves the performance of the two feature sets. 
    \item $\#e^{-}_O$/$\#e^{-}_{tot}$
\\* The product of the number of electrons in a single O atom and the number of O atoms in the compound of interest, divided by the total number of electrons in the compound of interest.  This is a global feature. This feature was introduced in the work by Wan et al. \cite{Wan2021}, where more details can be found.
\end{itemize}

\subsection{Models Performance}\label{Models}
Having introduced numerous descriptors of the oxygen vacancy formation energy in the previous section, we now test and compare the performance achieved by machine learning algorithms using different combinations of such descriptors. The machine learning algorithm of choice for most of the analysis is a random forest regressor \cite{Breiman2001}, both due to its reportedly solid performance across a variety of tasks, robustness against overfitting, speed and interpretability, and since, as we will show, it yields lower error than other tested regressors (kernel ridge, support vector machine, and linear regression). To give a more complete insight, we also examine correlation coefficients between the different descriptors.

\begin{figure}[!ht]
\includegraphics[width=8.3cm]{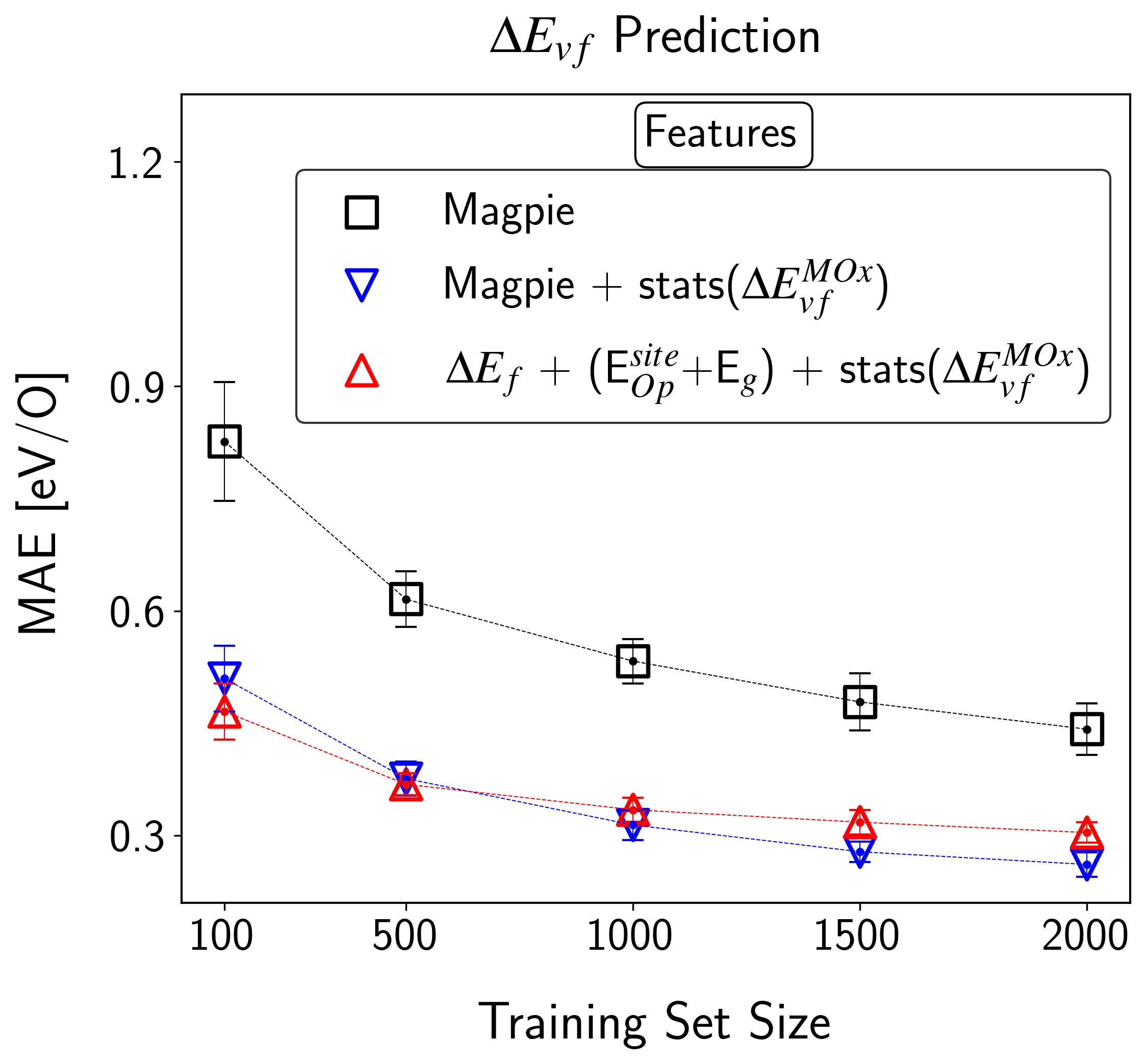}
	\caption{Mean absolute error upon prediction of $\Delta E_{vf}$ achieved by random forest models utilizing different feature sets as a function of training set size. The datapoints are centered on  averages over 10 random choices of training data, with error bars representing standard deviations. The marker indicate different feature sets. The black square indicates the Magpie feature set. The blue triangle indicates the feature set composed of Magpie features and information on the oxygen vacancy formation energy of binary oxides with the same cations as the ones neighboring the vacancy, i.e. Feature set I described in Section \ref{Features}. The red triangle indicates the feature set which uses information from DFT bulk calculations, i.e. Feature set II described in Section \ref{Features}. The test set size is always 500. Train and test set are randomly selected with the enforcement of the condition that each composition can only appear in one of the two sets. Without this condition, as many compounds have multiple non-equivalent vacancy sites, it is7 likely that a purely random train-test separation would lead to different sites of the same compound appearing in the two sets. }
\label{performance_this_work}
\end{figure}

We first test the predictive accuracy of the feature sets introduced in this work with different amounts of training data. Figure \ref{performance_this_work} displays the mean absolute error upon prediction of $\Delta E_{vf}$ achieved by a random forest model utilizing the two feature sets against training set size, with a fixed test set size of 500 datapoints. In all cases, training and testing set are selected such that, even in the presence of multiple different vacancy sites for the same compound, the sites are separated so that each compound only appears in one of the two sets, i.e. performance testing is only on entirely unseen compounds. For all ranges of training data, the two feature sets display a very similar performance, achieving a mean absolute error as low as 0.3 eV/O when the full training set size is utilized. We also compare the two feature sets with the Magpie feature set (black square) and highlight the doubling in accuracy (halving in MAE) that occurs when introducing the binary $\Delta E_{vf}$-related features (blue triangle). In addition to the performance tests just described, we also investigate the accuracy achieved when predicting the unrelaxed  and relaxation contributions to the vacancy formation energy (respectively, $\Delta E_{vf}^{UN}$ and $\Delta E_{vf}^{R}$, see Methods and Ref \cite{my_perovskite_paper}) independently from each other. We find that, when utilizing a training set of 2000 datapoints and Feature set I, the MAE upon prediction of $\Delta E_{vf}^{UN}$ is of $\sim$0.2 eV/O and that of $\Delta E_{vf}^{R}$ is of $\sim$0.1 eV/O. To put the above results in context, we note that, for the compounds included in our dataset, $\Delta E_{vf}^{UN}$ spans a range of $\sim$8eV, while $\Delta E_{vf}^{R}$ lies between 0 and -2eV for the vast majority of the cases, with a magnitude smaller than 1eV in over half of the cases (see Fig S3 and S4). This highlights the opportunity for achieving a more accurate $\Delta E_{vf}$ prediction by identifying features capable of capturing the mechanisms involved ionic relaxation upon vacancy formation (i.e. targeting the relaxation component), which can represent a development of interest for which the dataset presented in this work can be leveraged in the future. 

\begin{figure}[!ht]
\includegraphics[width=8.3cm]{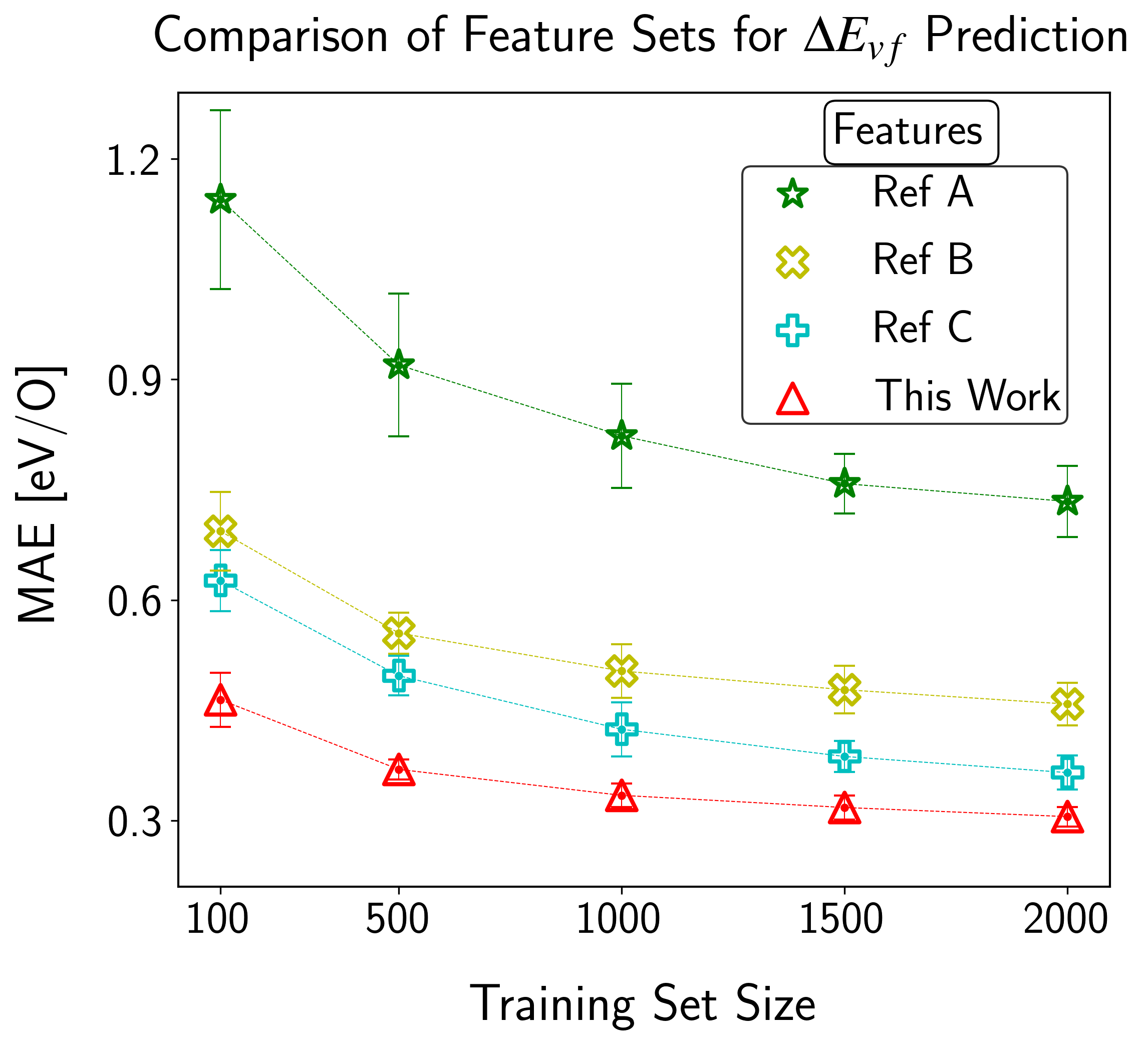}
	\caption{Mean absolute error of prediction of oxygen vacancy formation energy by random forest models utilizing different feature sets as a function of training set size. "Ref A" indicates the work by Wan et al. \cite{Wan2021} (where the set of features is: $\Delta\chi^{1stNN}$, $\#e^{-}_O$/$\#e^{-}_{tot}$, the total number of atoms, O atoms and electrons per formula unit, and the fraction of O atoms), "Ref B" the work by Deml et al \cite{Deml2015} (where the set of feature is: $\Delta E_f$, ($E_{Op}+0.75E_g$) and $\Delta\chi^{1stNN}$), "Ref C" the work by Wexler et al.\cite{Wexler2021} (where the set of features is: $V_r^{NN}$, $E_{g}$, $\Sigma E_b^{NN}$ and $\Delta E_{stab}$), and "This Work" indicates Feature set II described in the previous section. The test set size is always 500 and train and test set are randomly selected with the enforcement of the condition that each composition can only appear in one of the two sets.}
\label{performance_literature}
\end{figure}

We then compare the performance of the feature sets introduced in this work with those present in the literature. The differences in the training sets, both in terms of size and content, and in the regressors used in previous studies can lead to significant differences in performance upon extrapolation. Therefore, in order to achieve a transparent comparison, we opt for utilizing the same training set (the one presented in this work) and regressor (a random forest regressor) for different models that utilize the feature sets introduced in each work we consider. The results are displayed in Figure \ref{performance_literature}, where "Ref A" indicates the work by Wan et al. \cite{Wan2021} (marked with green stars), "Ref B" the work by Deml et al \cite{Deml2015} (marked with yellow crosses), "Ref C" the work by Wexler et al.\cite{Wexler2021} (marked with cyan plus signs), and "This Work" refers to the feature set introduced in this work that leverages DFT bulk calculation information (i.e. the one previously referred to as Feature Set II, marked with red triangles in both \ref{performance_this_work} and Fig \ref{performance_literature}). Through the comparison we see that, for all training set sizes, the random forest model using the feature set introduced in this work systematically outperforms the other models. Among the other models, the one utilizing the feature set developed by Wexler et al. \cite{Wexler2021} also shows remarkable performance, reaching a mean absolute error below 0.4 eV/O. We attribute a significant role in such achievement to the presence of features which clearly capture the difference both in specie and in oxidation state of the metal cations neighboring the vacancy, which we have seen to have a primary impact on $\Delta E_{vf}$ in Section \ref{Data}. We deem the absence of such type of feature to be the main reason limiting the accuracy reached by the model utilizing the feature set introduced by Deml et al \cite{Deml2015}. Lastly, the feature set chosen by Wan et al \cite{Wan2021} appears to produce significantly less accurate predictions, although we note that the aim of the authors was to identify a particularly simple set of features and avoid overfitting. Since previous works have focused on compounds with a non-zero DFT bandgap, and $\sim$ 25\% of the compounds in our dataset have a zero OQMD-computed DFT bandgap, we also reproduce Figure \ref{performance_literature} limiting the training and testing set to the compounds with $E_g>0$. The results of this additional performance test, displayed in Fig S5, are largely similar to those presented in Figure \ref{performance_literature}, with evidence of a modest performance improvement for the models utilizing the feature sets introduced by Wexler et al and Wan et al, but the overall predictive hierarchy between feature sets is left unchanged. In addition to the above comparison of the predictive power of random forest models utilizing different feature sets in the literature, we also consider performance of the convolutional neural network introduced by Witman et al \cite{Witman2023}. The lowest mean absolute error upon prediction of oxygen vacancy formation energy with a compound-wise train/test split (i.e. with the same logic applied in this study as described in the previous paragraph) is reported to be 0.45 $\pm$ 0.12 eV/O, and to have been achieved with 10-fold cross validation and a training set of over 10$^3$ datapoints. We however note that the training set utilized by the authors is composed of a variety of vacancy types, and that cation vacancy formation energies are likely to be less informative for oxygen vacancy prediction. We therefore point out that out of the 1481 defect energies present in the entire dataset, only 795 are from oxygen vacancies.

\begin{figure*}[!ht]
\includegraphics[width=16.5cm]{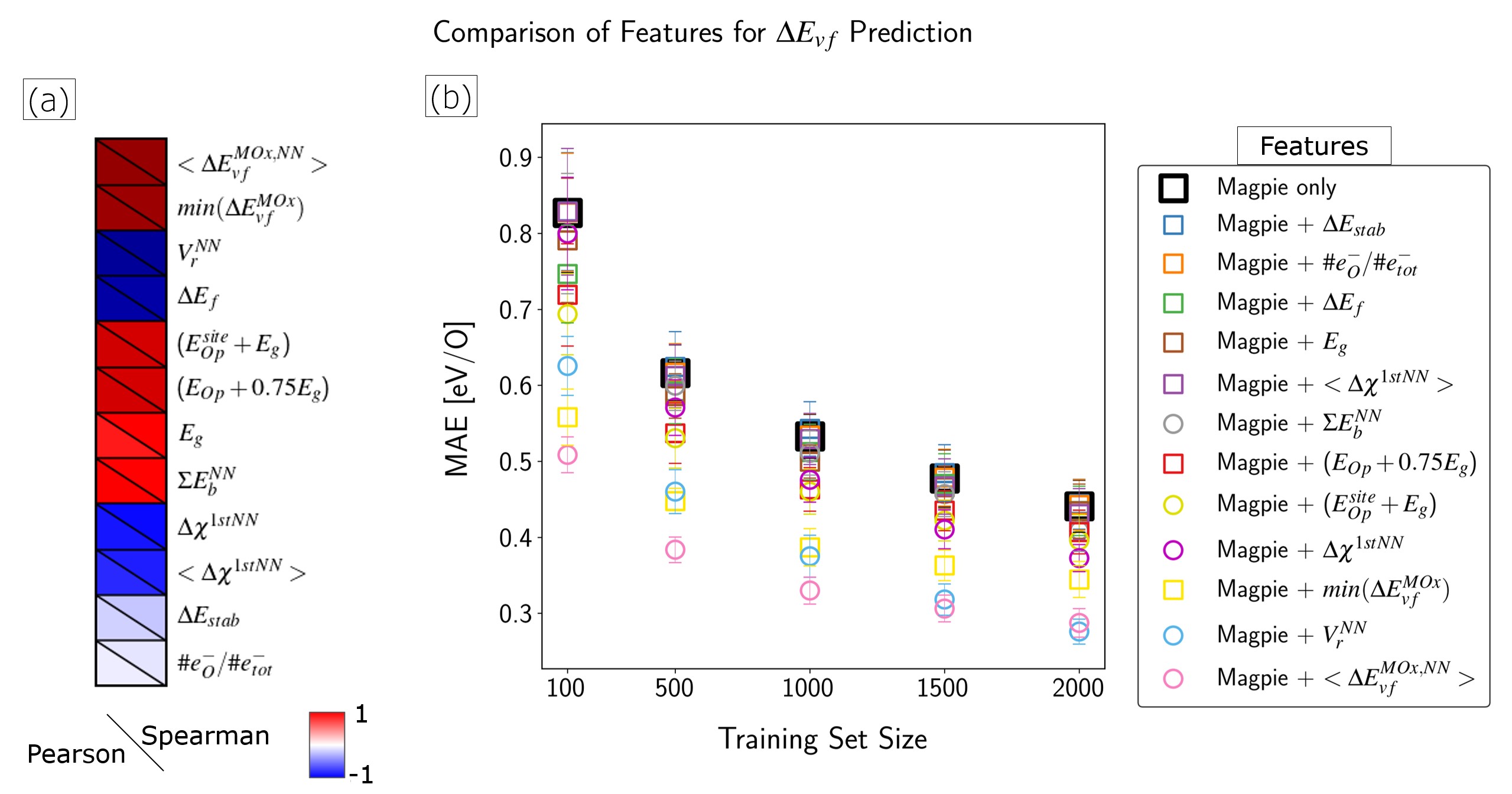}
	\caption{(a) Pearson (lower left triangle) and Spearman (upper right triangle) correlation coefficients of each feature with the oxygen vacancy formation energy. Red indicates positive correlation and blue indicates negative correlation, with darker shades indicating a stronger correlation. Feature names utilize the nomenclature introduced in Section \ref{Features} and are ordered top to bottom by magnitude (absolute value) of the correlation coefficients. (b) Mean absolute error of prediction of oxygen vacancy formation energy by random forest models utilizing different feature sets as a function of training set size. The black square indicates the case where only the Magpie feature set is utilized, while all other markers indicate the Magpie feature set plus one additional feature. When the additional feature is site-specific (i.e. contains information that depends on the site of the oxygen vacancy, such as nearest neighbors), the marker is circular, otherwise the marker is squared. The test set size is always 250 and train and test set are randomly selected with the enforcement of the condition that each composition can only appear in one of the two sets.}
\label{performance_features}
\end{figure*}

To gain further insight on the predictive utility of each of the individual features composing the feature sets just discussed, we now examine both their statistical correlation to $\Delta E_{vf}$, and the informational gain they carry in comparison with the Magpie feature set. Throughout this discussion we refer to the different features using the nomenclature introduced in Section \ref{Features}, where we explain in detail the significance and computation of each feature. In Figure \ref{performance_features} a we first report the Pearson and Spearman coefficients of each single feature in correlation with the vacancy formation energy. Features are ordered based on the magnitude of their correlation with the vacancy formation energy, and it can be clearly seen that features related to cation reduction ($\Delta E_{vf}^{MO_x}$, $\Delta E_{vf}^{MO_x,NN}$, and $V_r^{NN}$) exhibit the highest magnitude of correlation to the vacancy formation energy. The formation energy ($\Delta E_f$) and band structure-related properties containing information about the oxygen p-band ($E_{Op}^{site}+E_g$ and $E_{Op}+0.75E_g$) also show a strong correlation with $\Delta E_{vf}$, all having Pearson and Spearman coefficients above 0.67. In Figure \ref{performance_features} b we then examine the predictive accuracy of each feature when employed in a random forest model along with the Magpie feature set. The aim of this analysis is twofold. Firstly, to broaden the range of correlation types considered, as, while informational, the  Pearson and Spearman coefficients don't recognize all types of dependence between variables that can be leveraged by, eg. a random forest model. Secondly, to benchmark the utility of each additional feature compared to what could already be captured with an easily available, readily implementable, and widely used feature set. To perform the comparison, we compute the mean absolute error achieved by a random forest model with different training set sizes and a fixed test set size of 500 datapoints, starting with the Magpie feature set (black square), and then including one additional feature individually. We firstly notice that site-specific features, which are marked with circles, generally outperform global ones, marked with squares. A major exception to the above observation are the $\Delta E_{vf}^{MO_x}$-related features. In fact, in general throughout any training set size, and especially for lower training sizes, the features related to cation reduction are very apparently the best performing ones. Such features are the two introduced in this work: $\Delta E_{vf}^{MO_x}$ and $\Delta E_{vf}^{MO_x,NN}$, and the one introduced by Wexler et al \cite{Wexler2021}, $V_r^{NN}$. This evidence corroborates our earlier observation that the presence of element- and oxidation state- specific information in the feature set introduced by Wexler et al. plays a significant role in its competitive performance. We also remark that, while features such as the formation energy of the pristine structure $\Delta E_f$ do not exhibit any significant improvement in prediction accuracy of $\Delta E_{vf}$ when added to the Magpie feature set, this does not indicate a lack of utility in $\Delta E_{vf}$ prediction per se (as we can see by the correlation coefficients in Fig \ref{performance_features} a) but rather what is likely a redundancy with the information already captured by Magpie (Magpie is in fact utilized for predictions of $\Delta E_f$ itself \cite{Ward2016}\cite{Ward2017}\cite{Zhou2022}).

Finally, we test the effect that different regressors have on $\Delta E_{vf}$ predictions, when using the same training and feature set (see Fig S6). We test random forest (RF), support vector (SVR), kernel ridge (KRR) and linear regressors (LR), with the different sets of features examined in Figure \ref{performance_literature}. Random forest emerges as the best performing algorithm in all cases, with support vector and kernel ridge following, and linear regression achieving a significantly more limited accuracy. Furthermore, the feature set introduced in this work achieves the lowest mean absolute error across all regressors, its predictive power being by far the least affected by regressor choice among the feature sets considered. 

\subsection{New Predictions}\label{Predictions}
Having demonstrated the predictive power of random forest models utilizing the feature sets we introduced, we now exemplify their use in the search for new materials. Specifically, we choose to focus on new compounds for STCH applications. In this context, the perovskite family possesses a number of attractive properties, such as ability to withstand high temperature and to accomodate oxygen defects, and ease of oxygen diffusion. Having already studied both simple ABO$_3$ perovskites and B site ordered A$_2$BB$^{\prime}$O$_6$ double perovskites in Section \ref{Data}, we now take the next step in combinatorial complexity by considering two different cation species on both the A and the B site. In order to avoid the creation of large supercells to simulate disordered atomic configurations, we focus on ordered compounds. While numerous examples of B site ordered double perovskites are present in the literature, ordering on the A site appears to be more infrequent, often being stabilized by anion or cation vacancies \cite{King2010}. A number of examples of concurrent A and B site ordering in double perovskites have however been reported in compounds such as NaNdMnWO$_6$, where the A cations assume a layered-type of ordering, and the B cations assume a rocksalt configuration analogous to that of the double perovskites presented in Section \ref{Data} \cite{King2007}\cite{Knapp2006}. We therefore choose to focus on this structure type, and in particular the NaNdMnWO$_6$ structural prototype, which crystallizes in a monoclinic cell presenting both out-of-phase rotations of the oxygen octahedra about the [110] direction of the cubic structure and in-phase rotation of the octahedra about the [001] direction \cite{King2007}. 

To search for new STCH compounds with NaNdMnWO$_6$ structure type, we apply two screening criteria: thermodynamic stability and oxygen vacancy formation energy. In particular, we look for compounds within 25 meV/atom of the OQMD convex hull and with a $\Delta E_{vf}$ between 2 and 4.5 eV/O. In the past, this type of screening has been conducted by first computing the stability of all decorations of the structural prototype of interest, and subsequently computing the vacancy formation energy of the stable compounds identified in the previous step \cite{Emery2016}\cite{my_perovskite_paper}. In the present case, considering all distinct metal cations decorations of AA$^{\prime}$$_2$BB$^{\prime}$O$_6$ double perovskites' cation sites leads to millions of possible combinations.  In order to render the problem computationally tractable, we employ multiple strategies to decrease the number of candidates. 

As a first step, we reduce the number of compositions to be evaluated for thermodynamic stability. To do so, we first limit the possible cations occupying each site to specific groups that are more likely to lead to stable and ordered compounds \cite{Knapp2006}. In particular, we circumscribe the A site cations to alkali and alkali earth metals (K, Na, Rb, Ca, Sr, Ba), the A' ones to rare earth elements, the B cation to transition metals (any 3d, 4d, 5d transition metal), and the B' cation to early, high-valent transition metals (Ti, Zr, Hf, V, Nb, Ta, Cr, Mo, W). We then leverage the Feature set I introduced in Section \ref{Features} (the one containing only compositional and generic structural information) to predict the compounds' vacancy formation energies and apply a first, coarser, filter eliminating compounds having all predicted $\Delta E_{vf}$ values above 4.5 eV/O. We will eventually classify compounds with $\Delta E_{vf}$ in the 2-4.5 eV/O window as STCH candidates, therefore choosing a smaller higher bound than the one utilized in the high-throughput part of the work. The reason for this choice follows the observation that compounds with larger $\Delta E_{vf}$ would require particularly large values of the reduction entropy to be good STCH candidates \cite{my_perovskite_paper}, and, while all vacancy formation energies were calculated for the compounds in Section \ref{Data} and simply reporting the data comes at no additional cost, the present section aims to reduce as much as possible the number of vacancy calculations. Through the above described combined strategy, we reduce the number of compounds to $\sim$8600. We then perform stability calculations on the remaining compounds, and find over 600 of them to lie within 25 meV/atom of the OQMD convex hull. While the possibility of such compounds presenting cation disorder upon synthesis is conceivable, previous calculations on two recently experimentally reported mixed perovskites have shown that oxygen vacancy formation calculations performed on the ordered double perovskite equivalent of the mixed perovskite produced results accurate enough to guide candidate selection for STCH \cite{Qian2020}\cite{Qian2021}. 

Having identified several hundred new stable (within 25 meV/atom of the hull) AA$^{\prime}$$_2$BB$^{\prime}$O$_6$ candidates, we employ the random forest model using Feature set II introduced in Section \ref{Features} (the one containing information from the DFT calculation of the defect-free structure) to predict their oxygen vacancy formation energies. For each compound, six unique oxygen sites can be identified (one in each A layer, and four between the two A layers), making the prediction goal not only to differentiate $\Delta E_{vf}$ among different compounds, but also among different sites in the same compound, in order to predict the smallest value of $\Delta E_{vf}$ for each compound. In the majority of cases, the model predicts $\Delta E_{vf}$ of all the O sites in the same compound to be within 0.3 eV/O of each other. Given that all O sites neighbor both B cations (which are usually the ones where the majority of the charge localizes upon vacancy formation, as discussed in Section \ref{Data}), this observation is not surprising. It however indicates a potential obstacle to reliable ranking of $\Delta E_{vf}$ between defect sites as the mean absolute error achieved by the model is around the same size as the predicted differences between sites. We perform preliminary calculations on a subset of 5 compounds, and confirm the model's predictions of lack of large difference between vacancies before relaxation, but observe more significant differences upon relaxation in some cases, with the vacancies between the two A layers having a more negative $\Delta E_{vf}^{R}$ (and therefore a smaller $\Delta E_{vf}$). Based on these observations, we restrict our candidate search by considering only the O vacancy sites between the two A layers. We then select the compounds with the smallest predicted $\Delta E_{vf}$ lying in the target STCH range and directly compute their vacancy formation energy within the OQMD framework. Through such calculations, we both confirm the predictive power of the model, finding $\Delta E_{vf}$ values to lie in the exact range in over 75\% of the cases, and within 0.5 eV/O of the range in 99\% of the cases, and identify over 250 new STCH candidates, a complete list of which is reported in Table S3. Within such candidates, we highlight the presence of cations previously identified as promising for STCH applications: Mn$^{3+}$ (in compounds like (Ca,Sr)(Y,La,Ce,Pr,Sm,Tb,Gd,Dy)Mn(Ti,V)O$_6$), Mn$^{4+}$ (in compounds like Na(Y,Ce,Pr,Sm,Nd,Gd,Tb,Tm,Ho,Dy)Mn(Ti,Zr,Hf)O$_6$) and Co$^{3+}$ (in compounds like (Ca,Sr)(La,Nd,Pr,Dy)CoTiO$_6$).

\section{Conclusion}
The present work is dedicated to the DFT computation and machine learning prediction of oxygen vacancy formation energy in metal oxides. Firstly, we introduce and discuss the results from over $\sim$2500 direct calculations of $\Delta E_{vf}$ performed through the OQMD framework. We then leverage the dataset formed by such calculations, which contains over 60 different cation elements and 55 different structural spacegroups, to develop machine learning models of the vacancy formation energy. We introduce new descriptors, and test the performance of multiple regressors utilizing such descriptors, achieving a MAE upon prediction as low as 0.3 eV/O. We also investigate the comparative performance of other descriptors previously employed in the literature, and highlight the advantage of including cation and oxidation state -specific information. Finally, we utilize the models predictions to accelerate the search for STCH new compounds. We focus on A and B site ordered AA$^{\prime}$$_2$BB$^{\prime}$O$_6$ double perovskites, identify over 600 new stable compounds, and confirm over 250 of them to have $\Delta E_{vf}$ values appropriate for STCH applications.

\section{Acknowledgments}
This work was funded by the U.S. Department of Energy under Grant DE-EE0008089. S. G. acknowledges the Air Force Office of Scientific Research for support under Award No. FA9550-18-1-0136 (OQMD database). A. G. acknowledges the  Center for Hierarchical Materials Design (ChiMaD) under Award
No. 70NANB19H005 (ML models). A. S. C. and T. L. acknowledge funding from the Toyota Research Institute through the Accelerated Materials Design and Discovery program (ML representations). The high-throughput data was produced relying on the computing power provided by Quest high performance computing facility at Northwestern University.

\bibliography{mybibfile}

\clearpage
\onecolumngrid

 \begin{suppfigure}[!ht]
        \begin{center}
        {\mbox{\epsfig{file=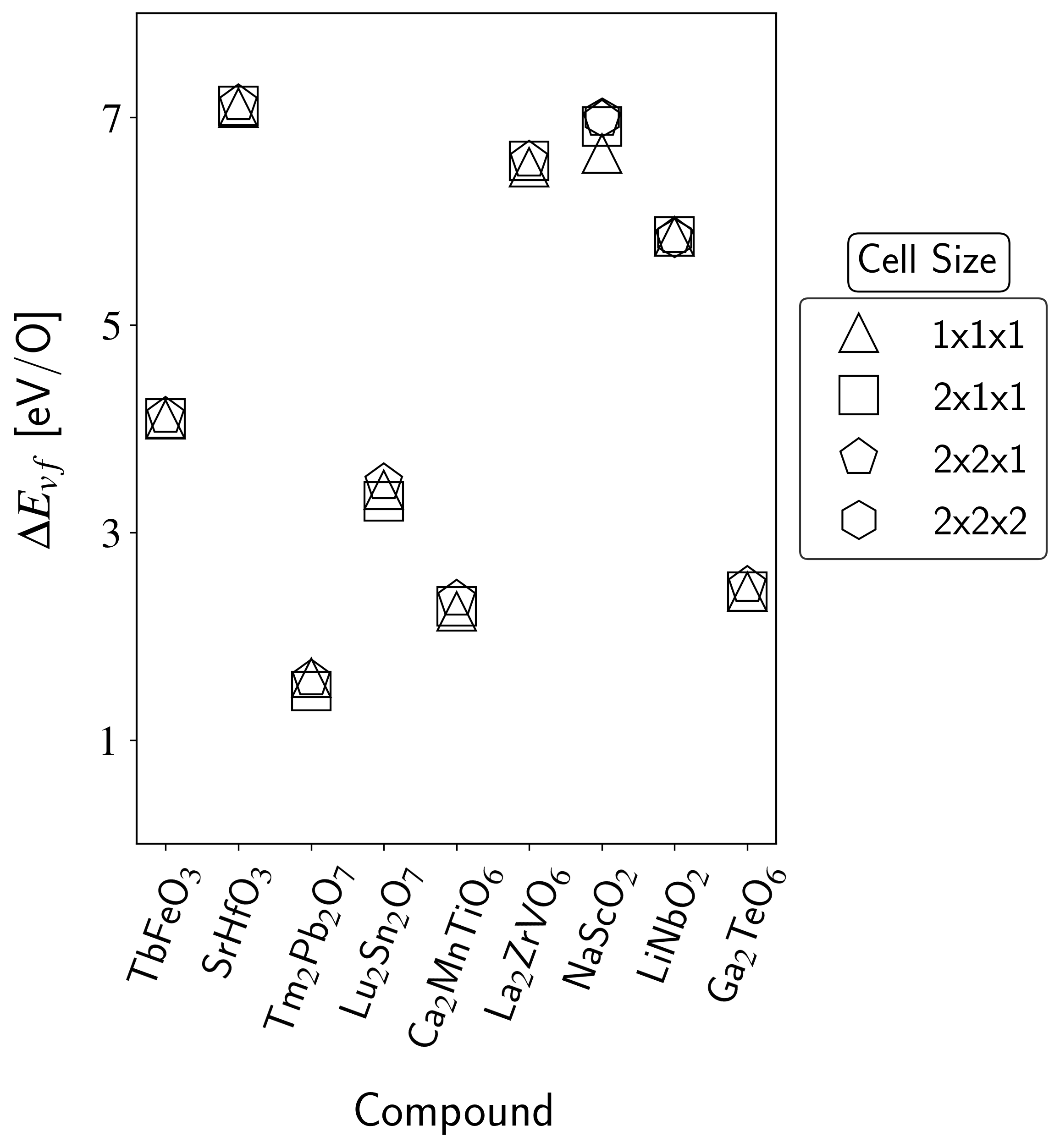,width=8.3cm}}}
        \caption{Change in oxygen vacancy formation energy of different compounds with size of the vacancy-containing cell. For all cell sizes greater than 16 atoms (which is the case for all compounds presented in the figure other than LiNbO$_2$, which has 8 atoms in its unit cell, and NaScO$_2$, which has 4) $\Delta E_{vf}$ is converged to within 0.1eV/O.}
        \label{Evf_binaries}
        \end{center}
        \renewcommand{\figurename}{S1}
    \end{suppfigure}

 \begin{suppfigure}[!ht]
        \begin{center}
        {\mbox{\epsfig{file=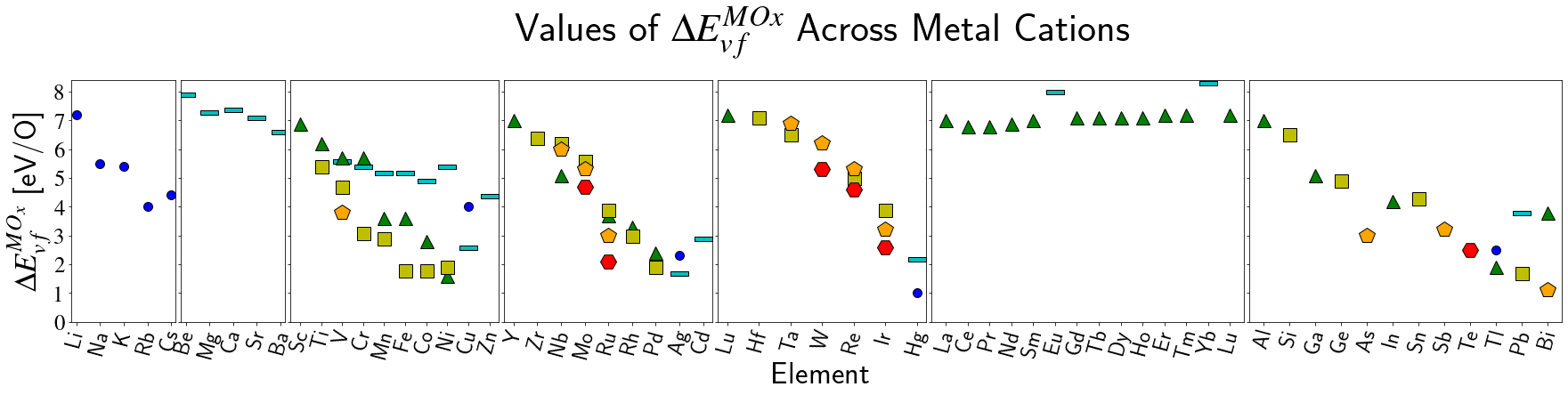,width=17.5cm}}}
        \caption{$\Delta E_{vf}^{MO_x}$ of all M$^{2x+}$ cations included in the dataset. Different markers indicate different oxidation states: blue circles for 1+, cyan lines for 2+, green triangles for 3+, yellow squares for 4+, orange pentagons for 5+, and red hexagons for 6+. Trends across groups, periods and oxidation states can be observed: rare-earth and alkali metal species have larger $\Delta E_{vf}^{MO_x}$ than most transition and post-transtion metals, and $\Delta E_{vf}^{MO_x}$ decreases with increasing period among the same transition metal series, and is larger for smaller oxidation states for the same metal.}
        \label{Evf_binaries}
        \end{center}
    \end{suppfigure}

    \newpage
    \begin{suppfigure}[!ht]
        \begin{center}
        {\mbox{\epsfig{file=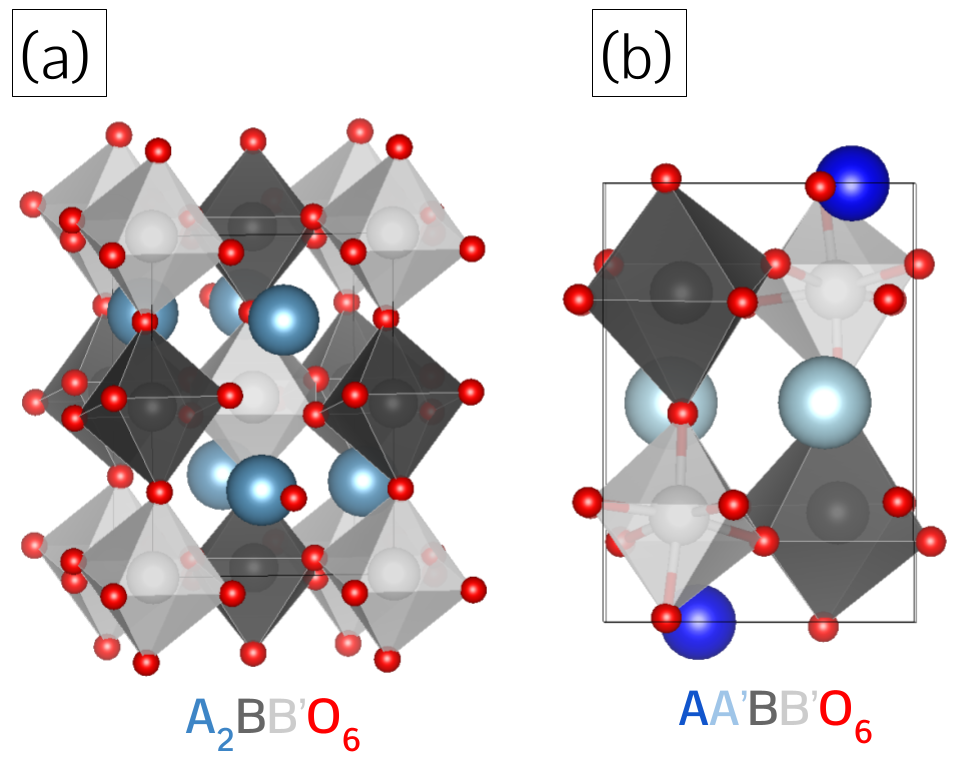,width=10.5cm}}}
        \caption{Visual representation of (a) the rock-salt type B site ordered A$_2$BB$^{\prime}$O$_6$ double perovskite structure, and (b) the A and B site ordered (with layered ordering on the A site and rock-salt type ordering on the B site) AA$^{\prime}$BB$^{\prime}$O$_6$ double perovskite structure.}
        \label{Evf_binaries}
        \end{center}
    \end{suppfigure}

 \begin{suppfigure}[!ht]
        \begin{center}
        {\mbox{\epsfig{file=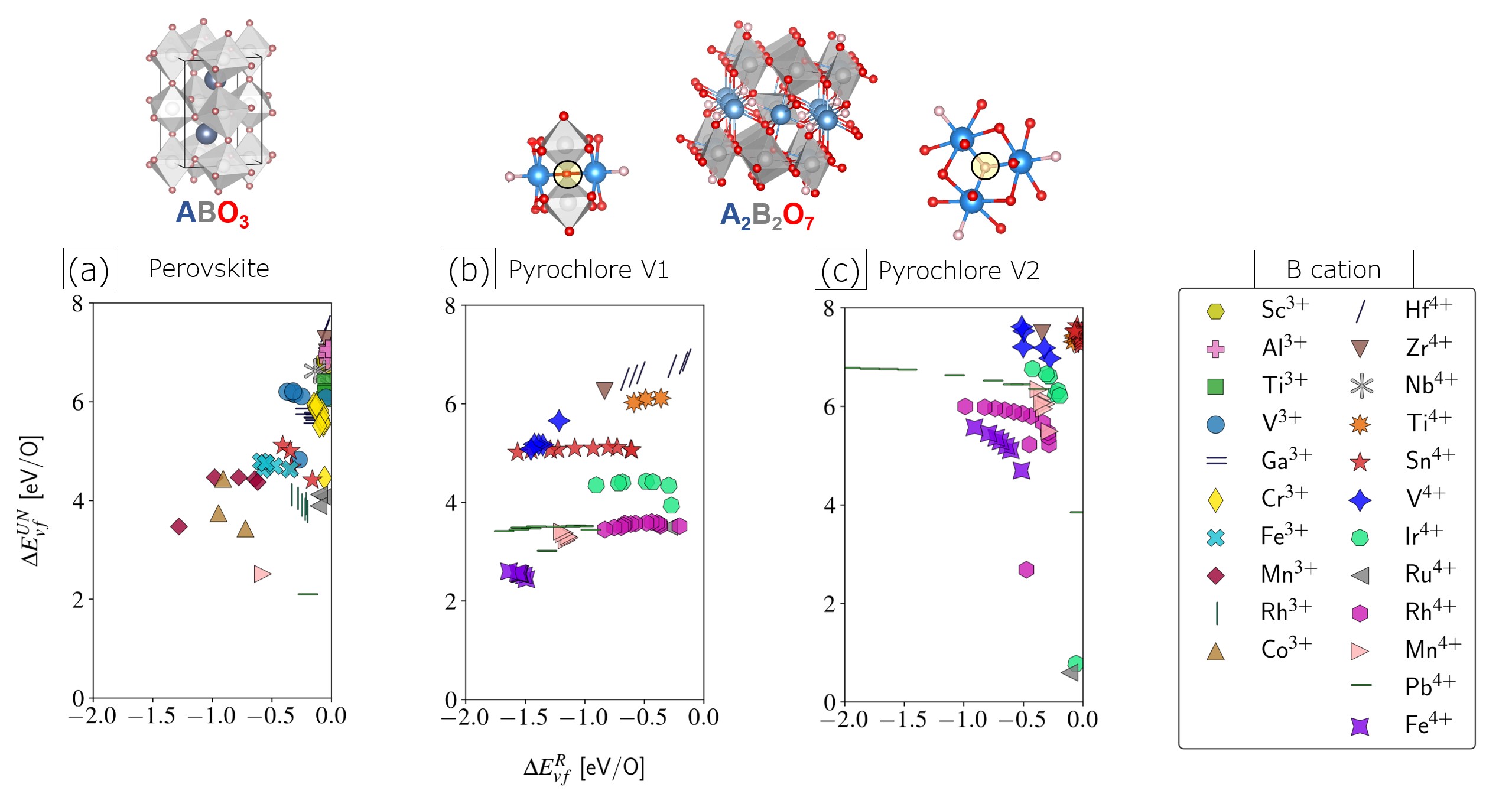,width=16cm}}}
        \caption{Unrelaxed  ($\Delta E_{vf,UN}$) vs relaxation ( $\Delta E_{vf,R}$) contributions to the oxygen vacancy formation energy of (a) ABO$_3$ perovskite oxides, and (b) and (c) A$_2$B$_2$O$_7$ pyrochlore oxides. In (a) $\Delta E_{vf}$ of the lowest energy vacancy is reported, as all vacancy sites have the same A and B cation nearest neighbors. In (b) $\Delta E_{vf}$ of sites with A and B cations (where B cations are the closets) as nearest neighbors is reporeted, and in (c) $\Delta E_{vf}$ of sites with only A cation nearest neighbors is reported. In all cases different markers and colors indicate the species and oxidation state of the B cation of each given compound.}
        \label{Evf_un_vs_rel}
        \end{center}
    \end{suppfigure}

 \begin{suppfigure}[!ht]
        \begin{center}
        {\mbox{\epsfig{file=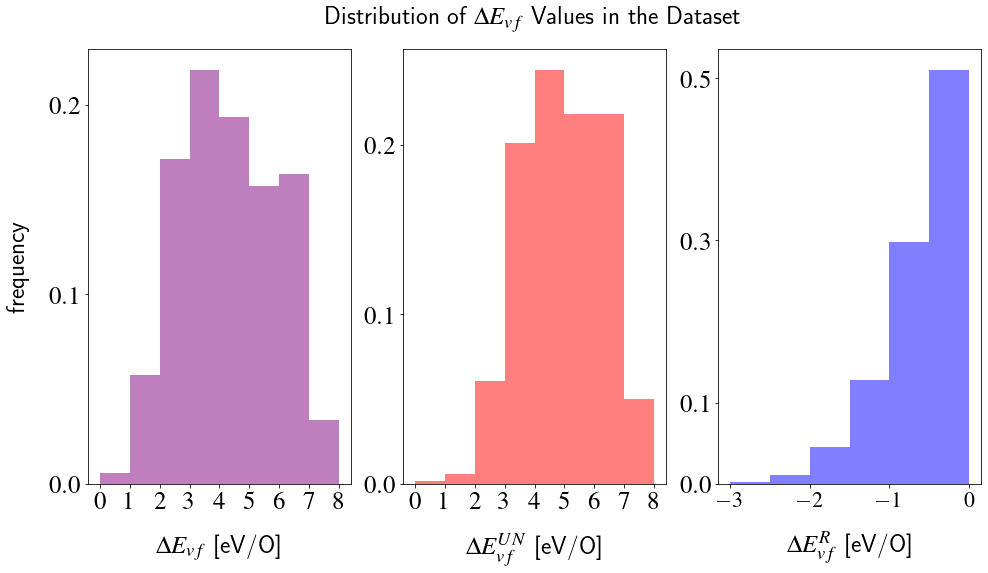,width=16cm}}}
        \caption{Distribution of values of the oxygen vacancy formation energy, and of its relaxation ($\Delta E_{vf,R}$) and unrelaxed ($\Delta E_{vf,UN}$) contributions for the compounds contained in the dataset.}
        \label{Evf_R_UN_frequency}
        \end{center}
    \end{suppfigure}
    
\newpage
  \begin{suppfigure}[!ht]
        \begin{center}
        {\mbox{\epsfig{file=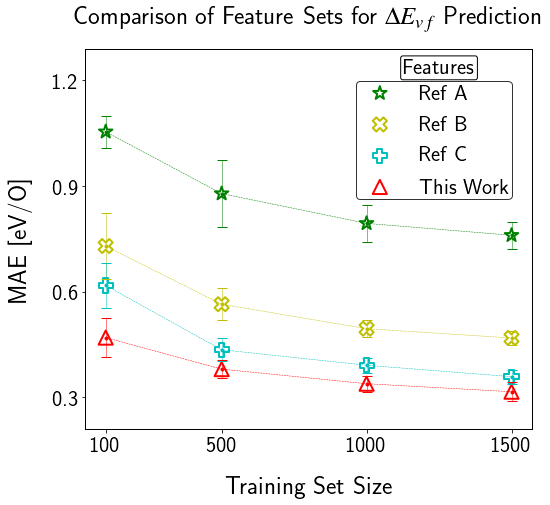,width=8.3cm}}}
        \caption{Mean absolute error of prediction of oxygen vacancy formation energy by random forest models utilizing different feature sets as a function of training set size. "Ref A" indicates the work by Wan et al. \cite{Wan2021} (where the set of features is: $\Delta\chi^{1stNN}$, $\#e^{-}_O$/$\#e^{-}_{tot}$, the total number of atoms, O atoms and electrons per formula unit, and the fraction of O atoms), "Ref B" the work by Deml et al \cite{Deml2015} (where the set of feature is: $\Delta E_f$, ($E_{Op}+0.75E_g$) and $\Delta\chi^{1stNN}$), "Ref C" the work by Wexler et al.\cite{Wexler2021} (where the set of features is: $V_r^{NN}$, $E_{g}$, $\Sigma E_b^{NN}$ and $\Delta E_{stab}$), and "This Work" indicates Feature Set B described in the previous section. The test set size is always 400 and train and test set are randomly selected from compounds in the dataset that have a non-zero OQMD computed DFT band gap, with the enforcement of the condition that each composition can only appear in one of the two sets.}
        \label{performance_literature_bgonly}
        \end{center}
    \end{suppfigure} 
    
  \begin{suppfigure}[!ht]
        \begin{center}
        {\mbox{\epsfig{file=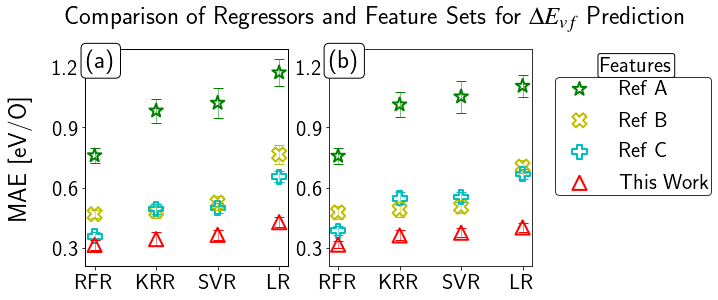,width=10.3cm}}}
        \caption{Mean absolute error of prediction of oxygen vacancy formation energy of different combinations of features and regressors. In (a) results are obtained by considering on the full dataset (2000 training data, 500 test),  and (b) only compounds with a non-zero OQMD computed DFT band gap (1500 training data, 400 test). RFR indicates random forest regression, SVR, support vector regression, KRR kernel ridge regression, and LR linear regression. "Ref A" indicates the work by Wan et al. \cite{Wan2021} (where the set of features is: $\Delta\chi^{1stNN}$, $\#e^{-}_O$/$\#e^{-}_{tot}$, the total number of atoms, O atoms and electrons per formula unit, and the fraction of O atoms), "Ref B" the work by Deml et al \cite{Deml2015} (where the set of feature is: $\Delta E_f$, ($E_{Op}+0.75E_g$) and $\Delta\chi^{1stNN}$), "Ref C" the work by Wexler et al.\cite{Wexler2021} (where the set of features is: $V_r^{NN}$, $E_{g}$, $\Sigma E_b^{NN}$ and $\Delta E_{stab}$), and "This Work" indicates Feature Set B described in the previous section.}
        \label{performance_regressors}
        \end{center}
    \end{suppfigure}

\newpage
\renewcommand\tablename{TABLE S}
\begin{longtable}{ p{.25\textwidth}| p{.25\textwidth}  } 
\caption{List of all DFT stable (25 meV/atom of the convex hull) A$_2$B$_2$O$_7$ pyrochlore compounds with a vacancy formation energy (of the lowest energy O site) within the window of interest for STCH applications. } \\

\hline
Composition   & $\Delta E_{vf}$ [eV/O]  \\

\midrule
\hline
Gd$_2$Pb$_2$O$_7$ & 2.0 \\
Sm$_2$Mn$_2$O$_7$ & 2.1 \\
Ca$_2$Ir$_2$O$_7$ & 2.1 \\
La$_2$Pd$_2$O$_7$ & 2.1 \\
Tm$_2$Pd$_2$O$_7$ & 2.1 \\
Er$_2$Pd$_2$O$_7$ & 2.1 \\
Ho$_2$Pd$_2$O$_7$ & 2.2 \\
Gd$_2$Mn$_2$O$_7$ & 2.2 \\
Tb$_2$Mn$_2$O$_7$ & 2.2 \\
Ce$_2$Pd$_2$O$_7$ & 2.2 \\
Dy$_2$Mn$_2$O$_7$ & 2.2 \\
Sm$_2$Pb$_2$O$_7$ & 2.2 \\
Dy$_2$Pd$_2$O$_7$ & 2.2 \\
Pr$_2$Pd$_2$O$_7$ & 2.2 \\
Tb$_2$Pd$_2$O$_7$ & 2.2 \\
Bi$_2$Rh$_2$O$_7$ & 2.2 \\
Ho$_2$Mn$_2$O$_7$ & 2.2 \\
Tm$_2$Mn$_2$O$_7$ & 2.2 \\
Nd$_2$Pd$_2$O$_7$ & 2.2 \\
Sm$_2$Pd$_2$O$_7$ & 2.2 \\
Nd$_2$Pb$_2$O$_7$ & 2.3 \\
Pr$_2$Pb$_2$O$_7$ & 2.4 \\
La$_2$Pb$_2$O$_7$ & 2.4 \\
Ce$_2$Pb$_2$O$_7$ & 2.4 \\
Tl$_2$Ge$_2$O$_7$ & 2.4 \\
Lu$_2$Rh$_2$O$_7$ & 2.6 \\
Tm$_2$Rh$_2$O$_7$ & 2.7 \\
Er$_2$Rh$_2$O$_7$ & 2.8 \\
Ho$_2$Rh$_2$O$_7$ & 2.9 \\
Dy$_2$Rh$_2$O$_7$ & 2.9 \\
Tb$_2$Rh$_2$O$_7$ & 2.9 \\
Gd$_2$Rh$_2$O$_7$ & 3.0 \\
Sm$_2$Rh$_2$O$_7$ & 3.1 \\
La$_2$Rh$_2$O$_7$ & 3.2 \\
Nd$_2$Rh$_2$O$_7$ & 3.2 \\
Ce$_2$Rh$_2$O$_7$ & 3.2 \\
Pr$_2$Rh$_2$O$_7$ & 3.2 \\
Tm$_2$Ir$_2$O$_7$ & 3.4 \\
Lu$_2$Sn$_2$O$_7$ & 3.4 \\
Tm$_2$Sn$_2$O$_7$ & 3.6 \\
Lu$_2$V$_2$O$_7$ & 3.6 \\
Tb$_2$Ir$_2$O$_7$ & 3.7 \\
Gd$_2$Ir$_2$O$_7$ & 3.7 \\
Tm$_2$V$_2$O$_7$ & 3.8 \\
Ho$_2$Sn$_2$O$_7$ & 3.8 \\
Er$_2$V$_2$O$_7$ & 3.8 \\
Ho$_2$V$_2$O$_7$ & 3.8 \\
Dy$_2$Sn$_2$O$_7$ & 3.8 \\
Nd$_2$Ir$_2$O$_7$ & 3.9 \\
Pr$_2$Ir$_2$O$_7$ & 4.0 \\
Gd$_2$Sn$_2$O$_7$ & 4.0 \\
La$_2$Ir$_2$O$_7$ & 4.1 \\
Sm$_2$Sn$_2$O$_7$ & 4.2 \\
Nd$_2$Sn$_2$O$_7$ & 4.3 \\
Pr$_2$Sn$_2$O$_7$ & 4.4 \\
Gd$_2$Fe$_2$O$_7$ & 4.4 \\
Dy$_2$V$_2$O$_7$ & 4.4 \\
La$_2$Sn$_2$O$_7$ & 4.5 \\
Ce$_2$Sn$_2$O$_7$ & 4.5 \\
Lu$_2$Ge$_2$O$_7$ & 4.8 \\
\end{longtable}
    
\renewcommand\tablename{TABLE S}
\begin{longtable}{ p{.25\textwidth} | p{.25\textwidth}  } 
\caption{List of all DFT stable (25 meV/atom of the convex hull) A$_2$BB$^{\prime}$O$_6$ double perovskite compounds with a vacancy formation energy (of the lowest energy O site) within the window of interest for STCH applications.} \\

\hline
Composition   & $\Delta E_{vf}$ [eV/O]  \\

\midrule
\hline
Ca$_2$ZnRuO$_6$ & 2.0 \\
Ba$_2$TlIrO$_6$ & 2.0 \\
Sr$_2$TiPdO$_6$ & 2.0 \\
Ca$_2$CoSbO$_6$ & 2.0 \\
Sr$_2$PdPbO$_6$ & 2.0 \\
Sr$_2$CoSbO$_6$ & 2.1 \\
Ba$_2$SnPbO$_6$ & 2.1 \\
Ba$_2$HgIrO$_6$ & 2.1 \\
Ba$_2$ZnRuO$_6$ & 2.1 \\
Sr$_2$TiMnO$_6$ & 2.1 \\
Ca$_2$TlSbO$_6$ & 2.2 \\
Ca$_2$InBiO$_6$ & 2.2 \\
Sr$_2$InBiO$_6$ & 2.2 \\
Ba$_2$TlSbO$_6$ & 2.2 \\
Sr$_2$HgIrO$_6$ & 2.2 \\
CdTePb$_2$O$_6$ & 2.2 \\
Sr$_2$FeRhO$_6$ & 2.2 \\
Sr$_2$SnPbO$_6$ & 2.2 \\
Ba$_2$HgTeO$_6$ & 2.2 \\
Sr$_2$NbCoO$_6$ & 2.3 \\
Ca$_2$NiRuO$_6$ & 2.3 \\
Sr$_2$NiRuO$_6$ & 2.3 \\
Sr$_2$RuPbO$_6$ & 2.3 \\
Sr$_2$TaTlO$_6$ & 2.3 \\
Ba$_2$ScBiO$_6$ & 2.3 \\
Ca$_2$ScBiO$_6$ & 2.3 \\
Ca$_2$TiMnO$_6$ & 2.3 \\
La$_2$GaCoO$_6$ & 2.3 \\
Sr$_2$TlSbO$_6$ & 2.3 \\
Ba$_2$CoRuO$_6$ & 2.3 \\
Ca$_2$TaCoO$_6$ & 2.3 \\
Sr$_2$ZrCrO$_6$ & 2.3 \\
Ca$_2$TlRuO$_6$ & 2.4 \\
Ba$_2$TaTlO$_6$ & 2.4 \\
Sr$_2$TlRuO$_6$ & 2.4 \\
LiLa$_2$BiO$_6$ & 2.4 \\
Sr$_2$MgRuO$_6$ & 2.4 \\
Sr$_2$TaCoO$_6$ & 2.4 \\
Sr$_2$CoRuO$_6$ & 2.4 \\
MnTePb$_2$O$_6$ & 2.4 \\
La$_2$CuPdO$_6$ & 2.4 \\
NaLa$_2$BiO$_6$ & 2.4 \\
Sr$_2$HfCrO$_6$ & 2.4 \\
Ca$_2$MgRuO$_6$ & 2.4 \\
Ca$_2$PdRuO$_6$ & 2.4 \\
Ba$_2$IrPbO$_6$ & 2.4 \\
La$_2$CdPdO$_6$ & 2.5 \\
Ca$_2$YBiO$_6$ & 2.5 \\
Sr$_2$ScBiO$_6$ & 2.5 \\
NbFePb$_2$O$_6$ & 2.5 \\
MnZn$_2$TeO$_6$ & 2.5 \\
Ba$_2$ZrPbO$_6$ & 2.5 \\
Sr$_2$SnPdO$_6$ & 2.5 \\
Ba$_2$MnSbO$_6$ & 2.6 \\
Sr$_2$HgTeO$_6$ & 2.6 \\
Sr$_2$ZrPbO$_6$ & 2.6 \\
Ba$_2$MgRuO$_6$ & 2.6 \\
Ba$_2$HfPbO$_6$ & 2.6 \\
La$_2$ZnPdO$_6$ & 2.6 \\
Sr$_2$TlIrO$_6$ & 2.6 \\
La$_2$CrNiO$_6$ & 2.7 \\
La$_2$CoPdO$_6$ & 2.7 \\
La$_2$NiPdO$_6$ & 2.7 \\
Ca$_2$NiIrO$_6$ & 2.7 \\
Ba$_2$YBiO$_6$ & 2.7 \\
Ba$_2$CuWO$_6$ & 2.7 \\
Sr$_2$HfPdO$_6$ & 2.7 \\
Ca$_2$CdIrO$_6$ & 2.7 \\
Sr$_2$MnIrO$_6$ & 2.8 \\
Ca$_2$ZrPdO$_6$ & 2.8 \\
La$_2$MgCrO$_6$ & 2.8 \\
Ca$_2$ZnIrO$_6$ & 2.8 \\
Ca$_2$TaMnO$_6$ & 2.8 \\
La$_2$MnCdO$_6$ & 2.8 \\
Sr$_2$YBiO$_6$ & 2.8 \\
Ba$_2$MnNbO$_6$ & 2.8 \\
Sr$_2$HfPbO$_6$ & 2.8 \\
Ca$_2$ZrMnO$_6$ & 2.8 \\
La$_2$MnZnO$_6$ & 2.8 \\
Sr$_2$ZnIrO$_6$ & 2.9 \\
Sr$_2$ZrMnO$_6$ & 2.9 \\
La$_2$CrCoO$_6$ & 2.9 \\
Ca$_2$MnNbO$_6$ & 2.9 \\
Ba$_2$TePbO$_6$ & 2.9 \\
Ca$_2$CoIrO$_6$ & 2.9 \\
La$_2$MgPdO$_6$ & 2.9 \\
Ba$_2$TaMnO$_6$ & 2.9 \\
Ca$_2$HfMnO$_6$ & 2.9 \\
Sr$_2$HfMnO$_6$ & 3.0 \\
Sr$_2$ZrPdO$_6$ & 3.0 \\
La$_2$CoRhO$_6$ & 3.0 \\
Sr$_2$CdIrO$_6$ & 3.0 \\
Ca$_2$VFeO$_6$ & 3.0 \\
Ca$_2$MgIrO$_6$ & 3.1 \\
Sr$_2$RuRhO$_6$ & 3.1 \\
Ba$_2$MnIrO$_6$ & 3.1 \\
Ba$_2$ZnIrO$_6$ & 3.1 \\
Sr$_2$TiRhO$_6$ & 3.1 \\
Ba$_2$CdTeO$_6$ & 3.1 \\
Sr$_2$FeTeO$_6$ & 3.2 \\
Sr$_2$NiIrO$_6$ & 3.2 \\
La$_2$AgIrO$_6$ & 3.2 \\
Ca$_2$FeTeO$_6$ & 3.2 \\
Sr$_2$BiRuO$_6$ & 3.2 \\
Sr$_2$CoTeO$_6$ & 3.2 \\
Ca$_2$CdTeO$_6$ & 3.2 \\
Sr$_2$CdTeO$_6$ & 3.2 \\
Ba$_2$MnTeO$_6$ & 3.2 \\
Sr$_2$NiTeO$_6$ & 3.2 \\
Ca$_2$MnSbO$_6$ & 3.2 \\
Sr$_2$TePbO$_6$ & 3.2 \\
Sr$_2$CoIrO$_6$ & 3.2 \\
Ba$_2$CdIrO$_6$ & 3.3 \\
Ba$_2$BiIrO$_6$ & 3.3 \\
Ca$_2$NiTeO$_6$ & 3.3 \\
La$_2$NiRhO$_6$ & 3.3 \\
Sr$_2$MnSbO$_6$ & 3.3 \\
Ca$_2$MnTeO$_6$ & 3.3 \\
Ca$_2$IrRhO$_6$ & 3.3 \\
Sr$_2$FeRuO$_6$ & 3.4 \\
Ba$_2$MgTeO$_6$ & 3.4 \\
Ca$_2$MnIrO$_6$ & 3.4 \\
La$_2$CuIrO$_6$ & 3.4 \\
Ca$_2$RuRhO$_6$ & 3.4 \\
Ca$_2$FeIrO$_6$ & 3.4 \\
Ba$_2$MgIrO$_6$ & 3.4 \\
TaInPb$_2$O$_6$ & 3.5 \\
Sr$_2$MgTeO$_6$ & 3.5 \\
Ca$_2$FeRuO$_6$ & 3.5 \\
Sr$_2$NbFeO$_6$ & 3.5 \\
Sr$_2$GaSbO$_6$ & 3.5 \\
Ba$_2$BiSbO$_6$ & 3.5 \\
Ca$_2$MgTeO$_6$ & 3.5 \\
Sr$_2$FeIrO$_6$ & 3.5 \\
Sr$_2$CrIrO$_6$ & 3.5 \\
La$_2$MnRhO$_6$ & 3.5 \\
ScZn$_2$SbO$_6$ & 3.5 \\
La$_2$CuRuO$_6$ & 3.6 \\
Ca$_2$CuReO$_6$ & 3.6 \\
Ba$_2$FeTeO$_6$ & 3.6 \\
CdRePb$_2$O$_6$ & 3.6 \\
Ba$_2$NbFeO$_6$ & 3.6 \\
Sr$_2$ZrRhO$_6$ & 3.6 \\
Ba$_2$InSbO$_6$ & 3.7 \\
Ba$_2$HfRhO$_6$ & 3.7 \\
Ca$_2$NbFeO$_6$ & 3.7 \\
Ca$_2$GaSbO$_6$ & 3.7 \\
Ca$_2$IrRuO$_6$ & 3.7 \\
Sr$_2$CrRuO$_6$ & 3.7 \\
Sr$_2$InIrO$_6$ & 3.7 \\
Ba$_2$TaFeO$_6$ & 3.7 \\
La$_2$MnCrO$_6$ & 3.7 \\
Ba$_2$InRuO$_6$ & 3.7 \\
Sr$_2$AlSbO$_6$ & 3.7 \\
Sr$_2$AlRuO$_6$ & 3.7 \\
Ba$_2$InIrO$_6$ & 3.8 \\
Sr$_2$AlIrO$_6$ & 3.8 \\
Sr$_2$HfRhO$_6$ & 3.8 \\
Ca$_2$InRuO$_6$ & 3.8 \\
Ca$_2$TiIrO$_6$ & 3.8 \\
Ba$_2$TiSnO$_6$ & 3.9 \\
Sr$_2$InRuO$_6$ & 3.9 \\
Sr$_2$TiVO$_6$ & 3.9 \\
Sr$_2$VSbO$_6$ & 3.9 \\
Ca$_2$CrRuO$_6$ & 3.9 \\
La$_2$FeRhO$_6$ & 3.9 \\
Ca$_2$AlSbO$_6$ & 3.9 \\
Ba$_2$ScRuO$_6$ & 3.9 \\
Ca$_2$AlRuO$_6$ & 3.9 \\
Sr$_2$BiSbO$_6$ & 3.9 \\
Sr$_2$InSbO$_6$ & 4.0 \\
Ba$_2$CdMoO$_6$ & 4.0 \\
La$_2$GaRhO$_6$ & 4.0 \\
Sr$_2$ScRuO$_6$ & 4.0 \\
Ca$_2$ScIrO$_6$ & 4.0 \\
Sr$_2$CuReO$_6$ & 4.0 \\
Sr$_2$TiIrO$_6$ & 4.0 \\
Ca$_2$NbRhO$_6$ & 4.0 \\
La$_2$MnFeO$_6$ & 4.0 \\
Ba$_2$NbRhO$_6$ & 4.1 \\
Ca$_2$ScRuO$_6$ & 4.1 \\
Sr$_2$TiSnO$_6$ & 4.1 \\
Sr$_2$CdMoO$_6$ & 4.1 \\
Ba$_2$ScSbO$_6$ & 4.1 \\
Sr$_2$CdReO$_6$ & 4.2 \\
La$_2$CdRuO$_6$ & 4.2 \\
Ca$_2$YRuO$_6$ & 4.2 \\
LiLa$_2$SbO$_6$ & 4.2 \\
Ca$_2$TiSnO$_6$ & 4.2 \\
NaLa$_2$SbO$_6$ & 4.2 \\
LiLa$_2$AsO$_6$ & 4.2 \\
Ba$_2$YRuO$_6$ & 4.2 \\
La$_2$MgIrO$_6$ & 4.3 \\
Sr$_2$NbRhO$_6$ & 4.3 \\
Sr$_2$YRuO$_6$ & 4.3 \\
Ca$_2$CdReO$_6$ & 4.3 \\
Ca$_2$ScSbO$_6$ & 4.3 \\
Sr$_2$ZrIrO$_6$ & 4.3 \\
Sr$_2$ScSbO$_6$ & 4.3 \\
Sr$_2$ZrVO$_6$ & 4.4 \\
La$_2$AlFeO$_6$ & 4.4 \\
La$_2$MnRuO$_6$ & 4.4 \\
La$_2$FeSnO$_6$ & 4.4 \\
NbCrPb$_2$O$_6$ & 4.4 \\
La$_2$ScRhO$_6$ & 4.4 \\
Ca$_2$ZrVO$_6$ & 4.4 \\
La$_2$FeRuO$_6$ & 4.5 \\
Sr$_2$TaRhO$_6$ & 4.5 \\
La$_2$NiSnO$_6$ & 4.5 \\
Ca$_2$YSbO$_6$ & 4.5 \\
Ba$_2$NbBiO$_6$ & 4.5 \\
Ba$_2$ZrIrO$_6$ & 4.5 \\
La$_2$MnSnO$_6$ & 4.5 \\
Ca$_2$HfVO$_6$ & 4.6 \\
Sr$_2$NbBiO$_6$ & 4.6 \\
Ba$_2$HfIrO$_6$ & 4.6 \\
La$_2$ScFeO$_6$ & 4.6 \\
La$_2$InGaO$_6$ & 4.6 \\
Ba$_2$ZrSnO$_6$ & 4.6 \\
Sr$_2$HfIrO$_6$ & 4.6 \\
La$_2$InFeO$_6$ & 4.6 \\
Sr$_2$YSbO$_6$ & 4.6 \\
La$_2$MgRuO$_6$ & 4.7 \\
Ca$_2$TaBiO$_6$ & 4.7 \\
Ca$_2$TiGeO$_6$ & 4.7 \\
Ca$_2$ZnReO$_6$ & 4.7 \\
La$_2$VNiO$_6$ & 4.7 \\
Ba$_2$FeMoO$_6$ & 4.7 \\
Sr$_2$TaBiO$_6$ & 4.7 \\
Ba$_2$TaInO$_6$ & 4.7 \\
Ca$_2$ZrSnO$_6$ & 4.7 \\
NbVPb$_2$O$_6$ & 4.7 \\
La$_2$CrFeO$_6$ & 4.8 \\
Sr$_2$FeReO$_6$ & 4.8 \\
Ca$_2$TaInO$_6$ & 4.8 \\
Ba$_2$ZrRuO$_6$ & 4.8 \\
Ba$_2$ZnMoO$_6$ & 4.8 \\
Ba$_2$TaBiO$_6$ & 4.8 \\
Sr$_2$ZrSnO$_6$ & 4.8 \\
Ba$_2$CoMoO$_6$ & 4.8 \\
Ca$_2$FeMoO$_6$ & 4.8 \\
Ba$_2$HfRuO$_6$ & 4.8 \\
Sr$_2$ZrRuO$_6$ & 4.9 \\
Sr$_2$FeMoO$_6$ & 4.9 \\
La$_2$VInO$_6$ & 4.9 \\
Sr$_2$HfSnO$_6$ & 4.9 \\
La$_2$CrInO$_6$ & 4.9 \\
Ba$_2$MnMoO$_6$ & 4.9 \\
La$_2$NiGeO$_6$ & 4.9 \\
Sr$_2$MgMoO$_6$ & 5.0 \\
Ba$_2$CrMoO$_6$ & 5.0 \\
Sr$_2$NbRuO$_6$ & 5.0 \\
Ca$_2$ReNiO$_6$ & 5.0 \\
\end{longtable}

\renewcommand\tablename{TABLE S}
\begin{longtable}{ p{.25\textwidth}| p{.25\textwidth}  } 
\caption{List of all DFT stable (25 meV/atom of the hull)  AA$^{\prime}$BB$^{\prime}$O$_6$ compounds with a vacancy formation energy (of the lowest energy O site) within the window of interest for STCH applications.} \\

\hline
Composition   & $\Delta E_{vf}$ [eV/O]  \\

\midrule
\hline

NaYbTaPbO$_6$ & 2.0  \\
BaEuNbTlO$_6$ & 2.0  \\
NaEuMnNbO$_6$ & 2.0  \\
NaEuTaPbO$_6$ & 2.0  \\
NaTmTiMnO$_6$ & 2.0  \\
RbNdZrPbO$_6$ & 2.0  \\
NaHoTiMnO$_6$ & 2.0  \\
NaErTiMnO$_6$ & 2.0  \\
KDyHfPbO$_6$ & 2.0  \\
NaEuTaMnO$_6$ & 2.0  \\
RbCeTaTlO$_6$ & 2.0  \\
KDyZrPbO$_6$ & 2.0  \\
KCeHfMnO$_6$ & 2.0  \\
NaPrTiMnO$_6$ & 2.0  \\
NaCeTiPdO$_6$ & 2.0  \\
NaPrTiPdO$_6$ & 2.0  \\
KCeMnNbO$_6$ & 2.1  \\
NaNdTiMnO$_6$ & 2.1  \\
BaEuTaTlO$_6$ & 2.1  \\
KGdHfPbO$_6$ & 2.1  \\
RbCeZrPbO$_6$ & 2.1  \\
KGdZrPbO$_6$ & 2.1  \\
KHoHfPbO$_6$ & 2.1  \\
KYHfPbO$_6$ & 2.1  \\
NaSmTiPbO$_6$ & 2.1  \\
RbCeHfPbO$_6$ & 2.2  \\
NaGdTiPbO$_6$ & 2.2  \\
NaTbTiPbO$_6$ & 2.2  \\
NaHoTiPbO$_6$ & 2.2  \\
NaDyTiPbO$_6$ & 2.2  \\
KLaMnNbO$_6$ & 2.2  \\
NaTmTiPbO$_6$ & 2.2  \\
CaLaTiCoO$_6$ & 2.2  \\
NaTmZrPbO$_6$ & 2.2  \\
KNdZrPbO$_6$ & 2.2  \\
KLaTaMnO$_6$ & 2.3  \\
SrNdTiCoO$_6$ & 2.3  \\
NaErZrPbO$_6$ & 2.3  \\
NaHoZrPbO$_6$ & 2.3  \\
NaYZrPbO$_6$ & 2.3  \\
NaTbZrPbO$_6$ & 2.3  \\
NaDyZrPbO$_6$ & 2.3  \\
NaGdZrPbO$_6$ & 2.3  \\
KPrZrPdO$_6$ & 2.3  \\
KCeTaMnO$_6$ & 2.3  \\
NaSmZrPbO$_6$ & 2.3  \\
NaNdZrPbO$_6$ & 2.3  \\
KLaZrPdO$_6$ & 2.4  \\
KLaHfPdO$_6$ & 2.4  \\
KCeHfPbO$_6$ & 2.4  \\
NaDyHfPbO$_6$ & 2.4  \\
CaPrTiCoO$_6$ & 2.4  \\
NaYHfPbO$_6$ & 2.4  \\
KCeZrPdO$_6$ & 2.4  \\
NaNdZrMnO$_6$ & 2.4  \\
NaCeZrMnO$_6$ & 2.4  \\
KCeHfPdO$_6$ & 2.4  \\
NaTbHfPbO$_6$ & 2.4  \\
KLaHfPbO$_6$ & 2.4  \\
CaNdTiCoO$_6$ & 2.4  \\
NaSmHfPbO$_6$ & 2.4  \\
NaNdHfMnO$_6$ & 2.4  \\
KCeZrPbO$_6$ & 2.5  \\
CaNdZrTlO$_6$ & 2.5  \\
KPrHfPbO$_6$ & 2.6  \\
KHoZrPbO$_6$ & 2.6  \\
NaPrZrMnO$_6$ & 2.6  \\
KCeTaTlO$_6$ & 2.6  \\
NaSmZrMnO$_6$ & 2.6  \\
KPrTaTlO$_6$ & 2.6  \\
BaLaTiMnO$_6$ & 2.6  \\
KTbZrPbO$_6$ & 2.6  \\
RbPrZrPbO$_6$ & 2.7  \\
NaPrHfMnO$_6$ & 2.7  \\
NaYHfMnO$_6$ & 2.7  \\
NaSmTaMnO$_6$ & 2.7  \\
NaDyHfMnO$_6$ & 2.7  \\
NaCeTaMnO$_6$ & 2.7  \\
NaNdTaMnO$_6$ & 2.7  \\
NaCeHfMnO$_6$ & 2.7  \\
NaPrTaMnO$_6$ & 2.7  \\
NaGdHfMnO$_6$ & 2.7  \\
NaTbHfMnO$_6$ & 2.7  \\
KSmZrPbO$_6$ & 2.8  \\
BaEuTaMnO$_6$ & 2.8  \\
KPrZrPbO$_6$ & 2.8  \\
KTbHfPbO$_6$ & 2.8  \\
CaDyTiCoO$_6$ & 2.8  \\
RbCeNbFeO$_6$ & 2.8  \\
NaTmHfPbO$_6$ & 2.8  \\
KSmHfPbO$_6$ & 2.9  \\
RbPrTaFeO$_6$ & 2.9  \\
KNdHfPbO$_6$ & 2.9  \\
NaNdHfPbO$_6$ & 2.9  \\
NaHoHfPbO$_6$ & 2.9  \\
NaErHfPbO$_6$ & 2.9  \\
NaSmHfMnO$_6$ & 3.0  \\
SrPrTiMnO$_6$ & 3.1  \\
NaCeTaTlO$_6$ & 3.1  \\
SrLaTiMnO$_6$ & 3.1  \\
CaSmTiMnO$_6$ & 3.1  \\
NaNdTaTlO$_6$ & 3.1  \\
KCeNbFeO$_6$ & 3.1  \\
SrCeTiMnO$_6$ & 3.1  \\
CaNdTiMnO$_6$ & 3.2  \\
CaLaTiMnO$_6$ & 3.2  \\
CaPrTiMnO$_6$ & 3.2  \\
KGdTaFeO$_6$ & 3.2  \\
NaHoHfMnO$_6$ & 3.2  \\
BaPrTiFeO$_6$ & 3.3  \\
BaLaTiFeO$_6$ & 3.3  \\
SrCeHfTlO$_6$ & 3.3  \\
BaYbTaFeO$_6$ & 3.4  \\
BaCeTiFeO$_6$ & 3.4  \\
CaTbTiMnO$_6$ & 3.4  \\
CaCeMnVO$_6$ & 3.4  \\
CaGdTiMnO$_6$ & 3.4  \\
CaPrMnVO$_6$ & 3.4  \\
CaCeTiMnO$_6$ & 3.4  \\
SrHoTiFeO$_6$ & 3.4  \\
CaNdMnVO$_6$ & 3.5  \\
SrDyTiFeO$_6$ & 3.5  \\
SrTbTiFeO$_6$ & 3.5  \\
CaNdTaCuO$_6$ & 3.5  \\
SrGdTiFeO$_6$ & 3.5  \\
BaLaTiRhO$_6$ & 3.6  \\
BaPrTiRhO$_6$ & 3.6  \\
CaGdMnVO$_6$ & 3.6  \\
CaLuTiFeO$_6$ & 3.6  \\
BaCeTiRhO$_6$ & 3.6  \\
CaYMnVO$_6$ & 3.6  \\
SrSmTiFeO$_6$ & 3.6  \\
CaDyMnVO$_6$ & 3.7  \\
BaEuTaFeO$_6$ & 3.7  \\
CaTmTiFeO$_6$ & 3.7  \\
CaErTiFeO$_6$ & 3.7  \\
SrNdTiFeO$_6$ & 3.7  \\
CaHoTiFeO$_6$ & 3.7  \\
CaDyTiFeO$_6$ & 3.7  \\
CaTbTiFeO$_6$ & 3.7  \\
CaGdTiFeO$_6$ & 3.8  \\
NaCeTiIrO$_6$ & 3.8  \\
NaPrTiIrO$_6$ & 3.8  \\
SrPrTiFeO$_6$ & 3.8  \\
KLaTiRuO$_6$ & 3.8  \\
KCeTiRuO$_6$ & 3.8  \\
BaEuNbRhO$_6$ & 3.8  \\
CaNdTiFeO$_6$ & 3.8  \\
SrCeTiFeO$_6$ & 3.8  \\
CaCeTiFeO$_6$ & 3.8  \\
CaYVFeO$_6$ & 3.9  \\
CaYTiFeO$_6$ & 3.9  \\
KCeZrIrO$_6$ & 3.9  \\
CaLaTiFeO$_6$ & 3.9  \\
NaTbVSbO$_6$ & 3.9  \\
SrNdTiRhO$_6$ & 3.9  \\
RbEuZrSbO$_6$ & 3.9  \\
RbTmTiTeO$_6$ & 3.9  \\
RbHoTiTeO$_6$ & 4.0  \\
CsSmTiTeO$_6$ & 4.0  \\
RbLuTiTeO$_6$ & 4.0  \\
CaErTiRhO$_6$ & 4.0  \\
CaHoTiRhO$_6$ & 4.0  \\
RbNdTiTeO$_6$ & 4.0  \\
CsPrTiTeO$_6$ & 4.0  \\
NaDyTiRuO$_6$ & 4.0  \\
CaTbTiRhO$_6$ & 4.0  \\
NaTbTiRuO$_6$ & 4.0  \\
CaGdTiRhO$_6$ & 4.0  \\
KCeTaRhO$_6$ & 4.0  \\
SrLaTiRhO$_6$ & 4.0  \\
RbLaTiTeO$_6$ & 4.0  \\
CaLuTiRhO$_6$ & 4.0  \\
SrCeTiRhO$_6$ & 4.0  \\
KEuZrSbO$_6$ & 4.0  \\
KEuNbRuO$_6$ & 4.0  \\
NaSmTiRuO$_6$ & 4.0  \\
CaTmTiRhO$_6$ & 4.0  \\
CaCeTiRhO$_6$ & 4.0  \\
RbCeTiTeO$_6$ & 4.0  \\
CaSmTiRhO$_6$ & 4.0  \\
NaYbZrSbO$_6$ & 4.0  \\
CaLaTiRhO$_6$ & 4.1  \\
NaEuNbIrO$_6$ & 4.1  \\
RbEuHfSbO$_6$ & 4.1  \\
NaCeTiRuO$_6$ & 4.1  \\
SrPrTiRhO$_6$ & 4.1  \\
CaPrTiRhO$_6$ & 4.1  \\
RbErTiTeO$_6$ & 4.1  \\
KCeTiTeO$_6$ & 4.1  \\
CsHoTiTeO$_6$ & 4.1  \\
RbYTiTeO$_6$ & 4.1  \\
CsDyTiTeO$_6$ & 4.1  \\
SrNdVFeO$_6$ & 4.1  \\
RbGdTiTeO$_6$ & 4.1  \\
NaPrTiRuO$_6$ & 4.1  \\
CsGdTiTeO$_6$ & 4.1  \\
CaYTiRhO$_6$ & 4.1  \\
RbTbTiTeO$_6$ & 4.1  \\
KErTiTeO$_6$ & 4.1  \\
KTmTiTeO$_6$ & 4.1  \\
KHoTiTeO$_6$ & 4.1  \\
CaNdTiRhO$_6$ & 4.1  \\
KEuHfSbO$_6$ & 4.1  \\
KPrTiTeO$_6$ & 4.1  \\
KYTiTeO$_6$ & 4.1  \\
KSmTiTeO$_6$ & 4.2  \\
NaLuTiSnO$_6$ & 4.2  \\
KNdTiTeO$_6$ & 4.2  \\
RbDyTiTeO$_6$ & 4.2  \\
KDyTiTeO$_6$ & 4.2  \\
KTbTiTeO$_6$ & 4.2  \\
NaCeCdMoO$_6$ & 4.2  \\
KGdTiTeO$_6$ & 4.2  \\
SrCeVFeO$_6$ & 4.2  \\
RbCeCdWO$_6$ & 4.2  \\
CaLuVFeO$_6$ & 4.2  \\
NaCeCdWO$_6$ & 4.2  \\
SrSmTiRhO$_6$ & 4.3  \\
CaGdVFeO$_6$ & 4.3  \\
NaTmCdWO$_6$ & 4.3  \\
NaTbCdWO$_6$ & 4.3  \\
NaYCdWO$_6$ & 4.3  \\
CaHoVFeO$_6$ & 4.3  \\
CaPrVFeO$_6$ & 4.3  \\
CaSmVFeO$_6$ & 4.3  \\
KPrCdWO$_6$ & 4.3  \\
CaDyVFeO$_6$ & 4.3  \\
NaYbHfSbO$_6$ & 4.3  \\
KCeZrRuO$_6$ & 4.3  \\
KLaZrRuO$_6$ & 4.3  \\
CaDyTiRhO$_6$ & 4.3  \\
CsEuTaTeO$_6$ & 4.3  \\
CaErVFeO$_6$ & 4.3  \\
CaTmVFeO$_6$ & 4.3  \\
BaEuCdWO$_6$ & 4.3  \\
NaYTaFeO$_6$ & 4.4  \\
NaPrCdWO$_6$ & 4.4  \\
SrCeHfFeO$_6$ & 4.4  \\
KPrZrRuO$_6$ & 4.4  \\
NaEuNbRuO$_6$ & 4.4  \\
KCeCdMoO$_6$ & 4.4  \\
NaTmTaFeO$_6$ & 4.4  \\
NaGdTaFeO$_6$ & 4.4  \\
NaDyTaFeO$_6$ & 4.4  \\
SrCeZrRhO$_6$ & 4.4  \\
NaErTaFeO$_6$ & 4.4  \\
NaYbNbRuO$_6$ & 4.4  \\
NaNdCdWO$_6$ & 4.4  \\
NaHoTaFeO$_6$ & 4.4  \\
KYbHfSbO$_6$ & 4.4  \\
NaSmCdWO$_6$ & 4.4  \\
SrGdNbCdO$_6$ & 4.5  \\
NaLaCdWO$_6$ & 4.5  \\
CaHoNbCdO$_6$ & 4.5  \\
BaCeNbCdO$_6$ & 4.5  \\
CaDyNbCdO$_6$ & 4.5  \\
NaYbTaRuO$_6$ & 4.5  \\
BaNdTaCdO$_6$ & 4.5  \\
KCeCdWO$_6$ & 4.5  \\
CsYTiTeO$_6$ & 4.5  \\
RbEuTaTeO$_6$ & 4.5  \\
\end{longtable}

\renewcommand\tablename{TABLE S}
\begin{longtable}{ p{.25\textwidth}| p{.25\textwidth}  } 
\caption{Unrelaxed oxygen vacancy formation energy of ground state binary metal oxides} \\

\hline
Composition   & $\Delta E_{vf}^{MO_x}$ [eV/O]  \\

\midrule
\hline
Ag$_2$O & 2.3 \\
Cs$_2$O & 4.4 \\
Cu$_2$O & 4.0 \\
Hg$_2$O & 1.0 \\
K$_2$O & 5.4 \\
Li$_2$O & 7.2 \\
Na$_2$O & 5.5 \\
Rb$_2$O & 4.0 \\
Tl$_2$O & 2.5 \\
AgO & 1.6 \\
BaO & 6.5 \\
BeO & 7.8 \\
CaO & 7.3 \\
CdO & 2.8 \\
CoO & 4.8 \\
CrO & 5.3 \\
CuO & 2.5 \\
EuO & 7.9 \\
FeO & 5.1 \\
HgO & 2.1 \\
MgO & 7.2 \\
MnO & 5.1 \\
NiO & 5.3 \\
PbO & 3.7 \\
SrO & 7.0 \\
VO & 5.5 \\
YbO & 8.2 \\
ZnO & 4.3 \\
Al$_2$O$_3$ & 7.0 \\
Bi$_2$O$_3$ & 3.8 \\
Ce$_2$O$_3$ & 6.8 \\
Co$_2$O$_3$ & 2.8 \\
Cr$_2$O$_3$ & 5.7 \\
Dy$_2$O$_3$ & 7.1 \\
Er$_2$O$_3$ & 7.2 \\
Fe$_2$O$_3$ & 3.6 \\
Ga$_2$O$_3$ & 5.1 \\
Gd$_2$O$_3$ & 7.1 \\
Ho$_2$O$_3$ & 7.1 \\
In$_2$O$_3$ & 4.2 \\
La$_2$O$_3$ & 7.0 \\
Lu$_2$O$_3$ & 7.2 \\
Mn$_2$O$_3$ & 3.6 \\
Nb$_2$O$_3$ & 5.1 \\
Nd$_2$O$_3$ & 6.9 \\
Ni$_2$O$_3$ & 1.6 \\
Pd$_2$O$_3$ & 2.4 \\
Pr$_2$O$_3$ & 6.8 \\
Rh$_2$O$_3$ & 3.3 \\
Ru$_2$O$_3$ & 3.7 \\
Sc$_2$O$_3$ & 6.9 \\
Sm$_2$O$_3$ & 7.0 \\
Tb$_2$O$_3$ & 7.1 \\
Ti$_2$O$_3$ & 6.2 \\
Tl$_2$O$_3$ & 1.9 \\
Tm$_2$O$_3$ & 7.2 \\
V$_2$O$_3$ & 5.7 \\
Y$_2$O$_3$ & 7.0 \\
CoO$_2$ & 1.8 \\
CrO$_2$ & 3.1 \\
FeO$_2$ & 1.8 \\
GeO$_2$ & 4.9 \\
HfO$_2$ & 7.1 \\
IrO$_2$ & 3.9 \\
MnO$_2$ & 2.9 \\
MoO$_2$ & 5.6 \\
NbO$_2$ & 6.2 \\
NiO$_2$ & 1.9 \\
PbO$_2$ & 1.7 \\
PdO$_2$ & 1.9 \\
ReO$_2$ & 5.0 \\
RhO$_2$ & 3.0 \\
RuO$_2$ & 3.9 \\
SiO$_2$ & 6.5 \\
SnO$_2$ & 4.3 \\
TaO$_2$ & 6.5 \\
TiO$_2$ & 5.4 \\
VO$_2$ & 4.7 \\
ZrO$_2$ & 6.4 \\
As$_2$O$_5$ & 3.0 \\
Bi$_2$O$_5$ & 1.1 \\
Ir$_2$O$_5$ & 3.2 \\
Mo$_2$O$_5$ & 5.3 \\
Nb$_2$O$_5$ & 6.0 \\
Re$_2$O$_5$ & 5.3 \\
Ru$_2$O$_5$ & 3.0 \\
Sb$_2$O$_5$ & 3.2 \\
Ta$_2$O$_5$ & 6.9 \\
V$_2$O$_5$ & 3.8 \\
W$_2$O$_5$ & 6.2 \\
IrO$_3$ & 2.6 \\
MoO$_3$ & 4.7 \\
ReO$_3$ & 4.6 \\
RuO$_3$ & 2.1 \\
TeO$_3$ & 2.5 \\
WO$_3$ & 5.3 \\

\end{longtable}

\end{document}